\documentclass[journal]{IEEEtran}
\makeatletter
\def\ps@headings{%
\def\@oddhead{\mbox{}\scriptsize\rightmark \hfil \thepage}%
\def\@evenhead{\scriptsize\thepage \hfil \leftmark\mbox{}}%
\def\@oddfoot{}%
\def\@evenfoot{}}
\makeatother
\usepackage[english]{babel}
\usepackage{fancyhdr}
\usepackage{graphicx}
\usepackage{subfigure}
\usepackage{amsmath}
\usepackage{epstopdf}
\usepackage{booktabs}
\usepackage[english]{babel}
\usepackage{amssymb}
\usepackage{tipa}
\usepackage{lpic}
\usepackage{stfloats}
\usepackage{algorithmic}
\usepackage{url}
\usepackage{verbatim}
\usepackage{empheq}
\usepackage{cases}
\usepackage{multirow}
\usepackage{booktabs}
\usepackage{threeparttable}
\usepackage[ruled,vlined]{algorithm2e}

\usepackage[noadjust]{cite}
\usepackage{enumerate}
\usepackage{amsthm}

\newtheorem{myDef}{Definition}
\newtheorem{myProp}{Proposition}
\newtheorem{myOpe}{Operation}

\ifCLASSINFOpdf

\else

\fi

\begin{document}

\title{Spectrum Resource Management for Multi-Access Edge Computing in Autonomous Vehicular Networks}

\author{\IEEEauthorblockN{Haixia Peng,~\IEEEmembership{Student Member,~IEEE}, Qiang Ye,~\IEEEmembership{Member,~IEEE}, Xuemin (Sherman) Shen,~\IEEEmembership{Fellow,~IEEE}}

\thanks{
H. Peng, Q. Ye, and X. Shen are with the Department of Electrical and Computer Engineering, University of Waterloo, Waterloo, ON, Canada, N2L 3G1 (e-mail: \{h27peng, q6ye, sshen\} @uwaterloo.ca).
}}

\maketitle
\IEEEpeerreviewmaketitle

\begin{abstract}
In this paper, a dynamic spectrum management framework is proposed to improve spectrum resource utilization in a multi-access edge computing (MEC) in autonomous vehicular network (AVNET). To support the increasing data traffic and guarantee quality-of-service (QoS), spectrum slicing, spectrum allocating, and transmit power controlling are jointly considered. Accordingly, three non-convex network utility maximization problems are formulated to slice spectrum among BSs, allocate spectrum among autonomous vehicles (AVs) associated with a BS, and control transmit powers of BSs, respectively. Via linear programming relaxation and first-order Taylor series approximation, these problems are transformed into tractable forms and then are jointly solved through an alternate concave search (ACS) algorithm. As a result, optimal spectrum slicing ratios among BSs, optimal BS-vehicle association patterns, optimal fractions of spectrum resources allocated to AVs, and optimal transmit powers of BSs are obtained. Based on our simulation, a high aggregate network utility is achieved by the proposed spectrum management scheme compared with two existing schemes.
\end{abstract}


\IEEEpeerreviewmaketitle

\section{Introduction}
\label{sec:Intro}

Recent advances in automobiles and artificial intelligence technology are promoting the developing of autonomous vehicles (AVs), which are becoming a reality \cite{hussain2018autonomous} and are expected to be commercialized and appear on the roads in the coming years \cite{human2018li}. However, salient challenges in computing and communication remain to be addressed to support AV applications. From the computing perspective, various computing tasks need to be carried out on board for real-time environment sensing and driving decision making \cite{su2018distributed}.
Moreover, enabling cooperative driving among AVs, such as platoon-based driving \cite{li2018consensus, peng2017performance, peng2017resource} and convoy-based driving \cite{su2018distributed, peng2018vehicular}, also requires extra computing tasks. From the communication perspective, the vehicular network enables AVs to support vehicular safety and non-safety related applications \cite{hussain2018autonomous}, share inter-vehicle information, and provide high-definition (HD) maps \cite{sabuau2017optimal}. Also, cooperative driving requires inter-vehicle communications for sharing position, velocity, acceleration, and other cruise control information among AVs \cite{peng2017performance}. All these required information exchanges among AVs increase the communication data traffic and are with different quality-of-service (QoS) requirements.

Some achievements have been made to overcome the challenges in computing and communication in vehicular networks. Edge computing has been regarded as an effective technology to enhance computing and storing capabilities in vehicular networks while alleviating traffic load to the core network \cite{software2018zhang, yuan2018toward, hui2018content}. Via moving computing and storing resources to servers placed at the edge of the core network, vehicles can offload its computing tasks to edge servers. Another potential method to address the computing issue is collaborative computing among vehicles \cite{hou2016vehicular, feng2017ave, su2018distributed}. In the scenarios with light computing task load, the utilization of computing resources can be improved through offloading computing tasks to the adjacent vehicles with idle computing power \cite{su2018distributed}. To address the communication issues in vehicular networks, interworking of multiple wireless access technologies has been widely accepted, such as the interworking of cellular network and dedicated short-range communications (DSRC) technologies \cite{abboud2016interworking}. To simultaneously address both computing and communication issues in vehicular networks, multi-access edge computing (MEC)\footnote{In 2017, mobile edge computing has been renamed to multi-access edge computing by the European Telecommunication Standards Institute (ETSI) to better reflect the growing interest and requirements in edge computing from non-cellular operators. } has recently been considered in some existing works 
\cite{ETSI2018V2X, hu2018vehicular}.

Inspired by existing works, a new architecture combines MEC with network function virtualization (NFV) and addresses the challenges in computing and communication in autonomous vehicular networks (AVNETs) \cite{Peng2018}. Via the MEC technology, 1) AVs with limited computing/storing resources can offload the tasks requiring high computing and storing requirements to MEC servers, such that a shorter response delay can be guaranteed through avoiding the data transfer between the core network and MEC servers; 2) multiple types of access technologies are permitted, thus moving AVs can access MEC servers via different base stations (BSs), such as Wi-Fi access points (Wi-Fi APs), road-side units (RSUs), White-Fi infostations, and evolved NodeBs (eNBs). Moreover, enabling NFV control module at each MEC server \cite{quan2018software, herrera2016resource, luo2018sdnmac}, the computing/storing resources placed at MEC servers can be dynamically managed and various radio spectrum resources can be abstracted and sliced to the BSs and then be allocated to AVs by each BS.

Efficient management for computing, storing, and spectrum resources is of paramount importance for the MEC-based AVNET. However, it is challenging to simultaneously manage the three types of resources while guaranteeing the QoS requirements for different AV applications, especially in the scenario with a high AV density. In this paper, we focus on spectrum resource management which can be extended to multiple resource allocation as our future work. The main contributions of this work are summarized as follows:
\begin{enumerate}
\item By considering the tradeoff between spectrum resource utilization and inter-cell interference, we develop a dynamic two-tier spectrum management framework for the MEC-based AVNET, which can be easily extended to other heterogenous networks.
\item Leveraging logarithmic and linear utility functions, we formulate three aggregate network-wide utility maximization problems to fairly slice spectrum resources among BSs connected to the same MEC server, optimize BS-vehicle association patterns and resource allocation, and control the transmit power of BS.
\item Linear programming relaxation and first-order Taylor series approximation are used and an alternate concave search (ACS) algorithm is designed to jointly solve the three formulated optimization problems.
\end{enumerate}

The remainder of this paper is organized as follows. First, the MEC-based AVNET is introduced in Section \ref{sec:System_M}, followed by the dynamic spectrum management framework and the communication model. In Section \ref{sec:Res_Man}, we formulate three optimization problems to slice and allocate spectrum resource among BSs and AVs and control transmit powers of BSs. Then, the three problems are transformed to tractable problems in Section \ref{sec:Pro_Analy} and an ACS algorithm is proposed to jointly solve them. In Section \ref{sec:simmu}, extensive simulation results are presented to demonstrate the performance of the proposed spectrum management framework. Finally, we draw concluding remarks in Section \ref{sec:CONCLUSIONS}.

\section{System Model}
\label{sec:System_M}

In this section, we first present an MEC-based AVNET architecture and a dynamic spectrum management framework, and then describe the communication model under the considered AVNET.

\subsection{MEC-based AVNET architecture}
\label{subsec:MEC_AVNETs}

\begin{figure}[htbp]
\centering
\includegraphics[height=3 in]{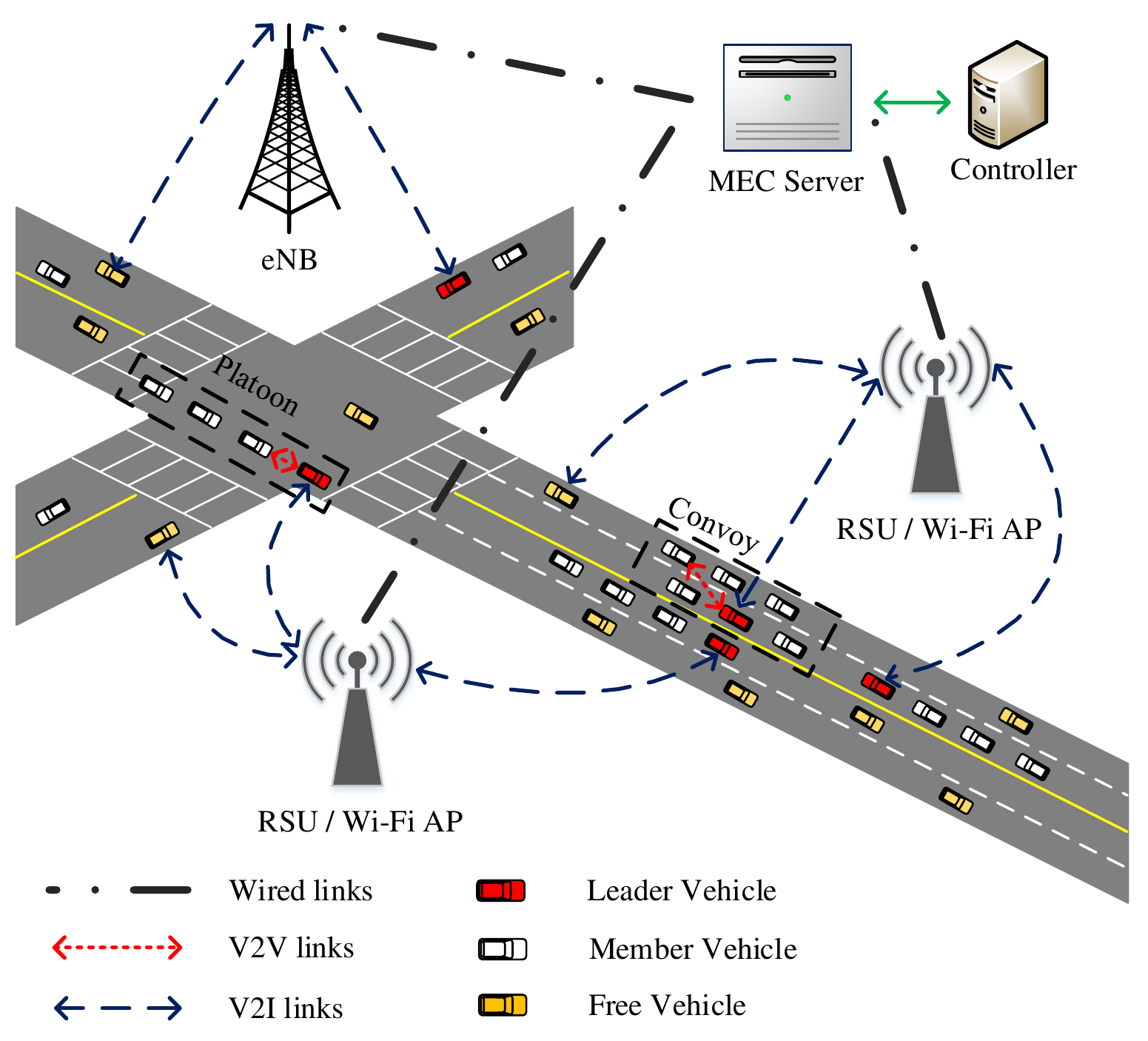}
\caption{An MEC-based AVNET model.}
\label{fig:systemTo}
\end{figure}

Based on a reference model suggested by the MEC ETSI industry specification group \cite{ETSI2018V2X}, we consider an MEC-based AVNET with one MEC server to support AV applications, as shown in Fig. \ref{fig:systemTo}. The MEC server allows AVs to access the edge computing/storing resources through different wireless access technologies.

To improve the cost efficiency of MEC server placement and provide short response delays to the AVs, the MEC server should be placed close to the edge of the core network rather than at each BS \cite{Peng2018} and the communication hops between an MEC server and an AV is assumed to be two. Thus, a large number of AVs within the coverages of several neighbored BSs can be served by the same MEC server and the enlarged service area of the MEC server can better overcome the challenges caused by high vehicle mobility. The total coverage area of BSs connected to an MEC server is defined as the service area of this server. To realize the resource virtualization process, including computing, storing, and spectrum resources, we consider a virtual wireless network controller at the MEC server. Through collecting information from the BSs and the AVs in the service area, resource management functions can run at the controller to adjust the virtual computing and storing resources to different AV tasks and to coordinate wireless access over the wide range of spectrum resources for AVs.

\subsection{Dynamic spectrum management framework}
\label{subsec:Dynamic_spec}

\begin{figure}[htbp]
\centering
\includegraphics[height=1.73 in]{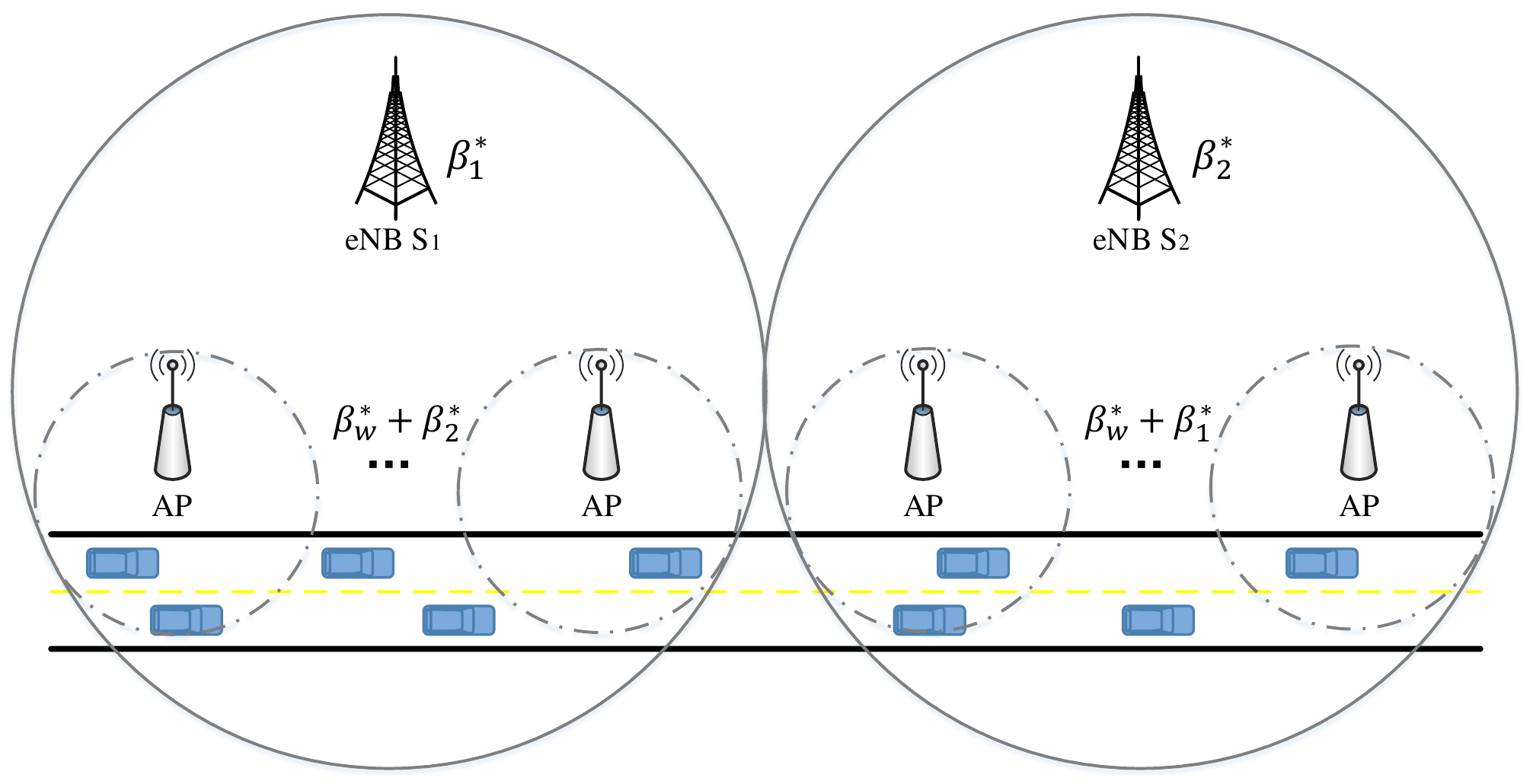}
\caption{A dynamic spectrum management framework.}
\label{fig:Framework}
\end{figure}

Due to the high vehicle mobility and heterogenous vehicular applications, AVNET topology and QoS requirements change frequently, and therefore, resource allocation should be adjusted accordingly. To improve spectrum resource utilization, a dynamic spectrum management framework is developed for downlink transmission. Taking a one-way straight road with two lanes as an example in Fig. \ref{fig:Framework}, two wireless access technologies, cellular and Wi-Fi/DSRC \cite{secinti2017software, liang2017vehicular}, are available to the AVs. Wi-Fi APs/RSUs and eNBs are uniformly deployed on one side of the road, where the $i$th Wi-Fi AP and the $j$th eNB are denoted by $W_i$ and $S_j$, respectively. The transmit power of each eNB, $P$, is fixed and high enough to guarantee a wide-area coverage, such that all AVs can receive sufficient strong control signal or information signal from eNBs. Denote $P_i^{\prime}$ as the transmit power of Wi-Fi AP $W_i$, which is lower than $P$ and dynamically adjusted by the controller. 
For AVs within the overlapping area of two BSs, only one of the BSs is associated for downlink transmission.

We divide the eNBs into two groups, denoted by $\mathcal{B}_1$ and $\mathcal{B}_2$, where eNBs in the same group are not neighbored to each other. ENBs $S_1$ and $S_2$ shown in Fig. \ref{fig:Framework} are the two target eNBs from the two different sets, where $S_1\in \mathcal{B}_1$ is adjacent to $S_2\in \mathcal{B}_2$. Set of Wi-Fi APs under the coverage area of eNB $S_j$ is denoted by $\mathcal{A}_j$. Denote the total available spectrum resource for vehicular applications to be $R^{\rm max}$. After collecting the application requests from AVs via BSs, the controller performs dynamic spectrum management for downlink transmission. The dynamic spectrum management procedure can be divided into two tiers as the following.
\begin{enumerate}
 \item \emph{Spectrum slicing among BSs:} The controller slices the spectrum resource, $R^{\rm max}$, into three slices with ratio set $\{\beta_1, \beta_2, \beta_w\}$ with $\beta_1+\beta_2+\beta_w=1$, and allocates them to eNBs in $\mathcal{B}_1$, eNBs in $\mathcal{B}_2$, and Wi-Fi APs, respectively.
\item \emph{Spectrum allocating among AVs:} Once the spectrum is sliced, each BS allocates its available spectrum resource to AVs associated to it. By allocating an appropriate amount of spectrum resources to each AV, the QoS requirements of various vehicular applications can be satisfied and the sum of transmission rates over the whole AVNET can be maximized.
\end{enumerate}
Spectrum slicing among BSs, spectrum allocating among AVs, and transmit power controlling for Wi-Fi APs are updated once the traffic load of each eNB fluctuates, which is in a relatively large time scale with respect to the dynamic communication environment. The traffic load of an eNB is defined as the average arrival traffic for AVs in the coverage area of the eNB.

\subsection{Communication model}
\label{subsec:Dynamic_spec}

Assume the three slices of spectrum resource are mutually orthogonal, therefore, there is no inter-tier interference. To improve the spectrum resource utilization, two levels of spectrum reusing are considered. The first level is reusing the spectrum resource $\beta_w R^{\rm max}$ among all the Wi-Fi APs as long as with an acceptable inter-cell interference. Moreover, we assume that the Wi-Fi APs with no overlapping coverage area with an eNB can reuse the spectrum allocated to that eNB. Thus, the interference to eNBs caused by the Wi-Fi APs can be controlled by adjusting the transmit powers of the Wi-Fi APs while the spectrum resource utilization can be further improved by allowing each Wi-Fi AP to reuse either the spectrum resource $(\beta_w + \beta_1) R^{\rm max}$ or $(\beta_w + \beta_2) R^{\rm max}$.

According to the dynamic spectrum management framework presented in Section \ref{subsec:Dynamic_spec}, all the eNBs in $\mathcal{B}_1$ reuse the spectrum resource $\beta_1 R^{\rm max}$ for downlink transmission. Denote $\mathcal{M}_j/M_j$ as the set/number of AVs within the coverage of eNB $S_j$. Then AV $k$, under the coverage of eNB $S_1$, i.e., $k\in \mathcal{M}_1$, experiences two kinds of interference to the corresponding downlink: from transmission of other eNBs in $\mathcal{B}_1$ and Wi-Fi APs in the coverage of eNBs in $\mathcal{B}_2$. Then, the spectrum efficiency at AV $k$ ($k\in \mathcal{M}_1$) from eNB $S_1$ can be expressed as
\begin{align}
\begin{split}
        \label{Spectrumeffi_S1}
         r^k_1 = {\rm log}_2 (1+
        \frac{P_1 G^k_1}{\sum \limits_{S_j\in\mathcal{B}_1,j\neq 1}P_j G^k_j + \sum \limits_{S_j\in\mathcal{B}_2}\sum \limits_{W_i\in\mathcal{A}_j} P_i^{\prime} G^{\prime k}_i + \sigma^2}),
\end{split}
\end{align}
where $G^k_j$ ($G^{\prime k}_i$) is the channel power gain between eNB $S_j$ (Wi-Fi AP $W_i$) and AV $k$, and $\sigma^2$ is the power spectrum density of the additive white Gaussian noise (AWGN). Similarly, the spectrum efficiency at AV $k$ ($k\in \mathcal{M}_2$) from eNB $S_2$, $r^k_2$, can be obtained. Let $R_j^k$ be the amount of spectrum allocated for AV $k$ from eNB $S_j$. Then, the achievable transmission rates of AV $k$ associated with eNBs $S_1$ (or $S_2$) can be expressed as
\begin{align}
        \label{Utility_S1}
        \gamma^k_1 = R_1^k r^k_1 \,\,\, ({\rm or} \, \gamma^k_2 = R^k_2 r^k_2).
\end{align}

Denote $\mathcal{N}_i/N_i$ as set/number of AVs within the coverage of Wi-Fi AP $W_i$. Let $R_{2,g}^{\prime k}$ and $R_{w,g}^{\prime k}$ be the amount of spectrum allocated to AV $k$ from $\beta_2 R^{\rm max}$ and $\beta_w R^{\rm max}$, respectively, by Wi-Fi AP $W_g$ under the coverage of eNB $S_1$ (i.e., $W_g \in \mathcal{A}_1$). Then the spectrum efficiencies at AV $k$ from Wi-Fi AP $W_g$ include the following two parts,
\begin{align}
\begin{split}
        \label{Spectrumeffi_Sg2}
        r_{2,g}^{\prime k} =  {\rm log}_2 (1+
        \frac{P_g^\prime G^{\prime k}_g}{\sum \limits_{W_i\in\mathcal{A}_1,i\neq g}P_i^{\prime} G^{\prime k}_i + \sum \limits_{S_j\in\mathcal{B}_2} P_j G^k_j + \sigma^2}) \\
        r_{w,g}^{\prime k} = {\rm log}_2 (1+
        \frac{P_g^\prime G^{\prime k}_g}{\sum \limits_{W_i\in\{\mathcal{A}_1\cup\mathcal{A}_2\},i\neq g}P_i^{\prime} G^{\prime k}_i + \sigma^2}). \\
\end{split}
\end{align}
And the achievable transmission rate of a tagged AV $k$ associated with Wi-Fi AP $W_g$, i.e., $k\in \cup_{W_g\in \mathcal{A}_1}\mathcal{N}_g$, can be expressed as
\begin{align}
\begin{split}
        \label{Spectrumeffi_Sg}
        \gamma^{\prime k}_g = & R_{2,g}^{\prime k} r_{2,g}^{\prime k} + R_{w,g}^{\prime k} r_{w,g}^{\prime k}.
\end{split}
\end{align}

Let $R_{1,h}^{\prime k}$ and $R_{w,h}^{\prime k}$ be the amount of spectrum allocated for AV $k$ from $\beta_1 R^{\rm max}$ and $\beta_w R^{\rm max}$, respectively, by Wi-Fi AP $W_h$ under the coverage of eNB $S_2$ (i.e., $W_h\in \mathcal{A}_2$), and $r_{1,h}^{\prime k}$ and $r_{w,h}^{\prime k}$ be the spectrum efficiencies at AV $k$ from Wi-Fi AP $W_h$. Similarly, the achievable transmission rate of a tagged AV $k$ associated with Wi-Fi AP $W_h$, i.e., $k\in \cup_{W_h\in \mathcal{A}_2}\mathcal{N}_h$, can be given by 
\begin{align}
\begin{split}
        \label{Spectrumeffi_Sh}
        \gamma^{\prime k}_h = R_{1,h}^{\prime k} r_{1,h}^{\prime k} + R_{w,h}^{\prime k} r_{w,h}^{\prime k}.
\end{split}
\end{align}

\section{Resource Management Scheme}
\label{sec:Res_Man}

We consider two kinds of traffic for each AV: delay-sensitive traffic and delay-tolerant traffic. Examples of AV's delay-sensitive traffic include rear-end collision avoidance and platooning/convoying. The delay-tolerant traffic can be HD map information downloading and infotainment services. Denote $p$ as the probability that an AV generates a delay-sensitive request. To accommodate the large amounts of data traffic generated by AVs while guaranteeing different QoS requirements for different applications, efficient resource management schemes are very important.

For downlink transmission to accommodate AVs' delay-sensitive requests, the transmission delay from eNB $S_j$ or Wi-Fi AP $W_i$ should be guaranteed statically. Let $L_s$ and $\lambda_s$ be the size and the arrival rate of the delay-sensitive packet. From \cite{Ye2018}, the maximum delay requirement, $D_{\rm max}$, can be transformed to a lower bound of the required transmission rate to guarantee that the downlink transmission delay exceeding $D_{\rm max}$ at most with probability $\varrho$, which can be expressed as
\begin{align}
        \label{minrate}
        \gamma_{\rm min} = -\frac{L_s {\rm log}\varrho}{D_{\rm max}{\rm log}(1-{\rm log}\varrho/(\lambda_s D_{\rm max}))}.
\end{align}

\subsection{Spectrum Resource Allocation}
\label{subsec:Spe_Re_Allo}

To address complicated resource allocation, we will introduce a two-tier approach, including spectrum slicing among BSs and spectrum allocating among AVs, as following.

\textbf{Spectrum slicing among BSs:} Based on the dynamic spectrum management framework, the total available spectrum resources are sliced or divided according to the ratio set $\{\beta_1, \beta_2, \beta_w\}$ for different BSs. The main concern for spectrum slicing is fairness among BSs. To this end, a logarithmic utility function, which is concave and with diminishing marginal utility \cite{Ye2018}, is considered to achieve a certain level of fairness among BSs \cite{ye2013user, liang2016virtual}.

For AV $k$ within the coverages of Wi-Fi APs, binary variables $x_j^k$ and $x_i^{\prime k}$ represent the BS-vehicle association patterns, where $x_j^k=1$ (or $x_i^{\prime k}=1$) means AV $k$ is associated with eNB $S_j$ (or Wi-Fi AP $W_i$), $x_j^k=0$ (or $x_i^{\prime k}=0$) otherwise. Denote $\overline{\mathcal{M}}_j/\overline{M}_j$ as set/number of AVs within the coverage of eNB $S_j$ while outside of Wi-Fi APs. Then, the utility for vehicle $k$ associated to eNBs or Wi-Fi APs is
\begin{align}
\label{scheme1}
u_k=
\begin{cases}
u^k_1 = {\rm log}(\gamma^k_1), &\mbox{if $k\in\overline{\mathcal{M}}_1\cup \{k|x_1^k=1\}$} \\
u^k_2 = {\rm log}(\gamma^k_2), &\mbox{if $k\in\overline{\mathcal{M}}_2\cup \{k|x_2^k=1\}$} \\
u^{\prime k}_g = {\rm log}(\gamma^{\prime k}_g), &\mbox{if $k\in\mathcal{N}_g\cap \{k|x_g^k=1\}$} \\
u^{\prime k}_h = {\rm log}(\gamma^{\prime k}_h), &\mbox{if $k\in\mathcal{N}_h\cap \{k|x_h^k=1\}$}.
\end{cases}
\end{align}

The aggregated network-wide utility is defined as the summation of utility of each individual AV. Let $\textbf{R}=\{R^k_1, R^k_2\}$ and $\textbf{R}^\prime=\{R_{2,g}^{\prime k}, R_{w,g}^{\prime k}, R_{1,h}^{\prime k}, R_{w,h}^{\prime k}\}$ be the matrices describing spectrum allocation among AVs by eNBs and by Wi-Fi APs, respectively. For given BS-vehicle association patterns with fixed transmit power of each Wi-Fi AP, the aggregated network-wide utility maximization problem can be given by
\begin{equation}
\begin{split}
\label{P1}
\mathbf{P1}: \max_{\substack{\beta_1, \beta_2, \beta_w \\\textbf{R}, \textbf{R}^\prime} } & \sum_{k\in \overline{\mathcal{M}}_1} u^k_{1} + \sum_{W_g} \sum_{k\in \mathcal{N}_g} (x^k_{1} u^k_{1} + x^{\prime k}_g u^{\prime k}_g) \\
                     & + \sum_{k\in \overline{\mathcal{M}}_2} u^k_{2} + \sum_{W_h} \sum_{k\in \mathcal{N}_h} (x^k_{2} u^k_{2} + x^{\prime k}_h u^{\prime k}_h) \\
\end{split}
\end{equation}
\addtocounter{equation}{-1}
\begin{subnumcases}
{\rm s.t.}
\beta_1, \beta_2, \beta_w \in [0,1] \label{BRP1_Cons_1} \\
\beta_1+\beta_2+\beta_w = 1 \label{BRP1_Cons_2} \\
\sum_{k\in \overline{\mathcal{M}}_1} R^k_1 + \sum_{W_g} \sum_{k\in\mathcal{N}_g} x^k_1 R^k_1 = \beta_1 R^{\rm max} \label{BRP1_Cons_3}\\
\sum_{k\in \overline{\mathcal{M}}_2} R^k_2 + \sum_{W_h} \sum_{k\in\mathcal{N}_h} x^k_2 R^k_2 = \beta_2 R^{\rm max}
\label{BRP1_Cons_4}\\
\sum_{k\in\mathcal{N}_g} x^{\prime k}_g R^{\prime k}_{l,g} = \beta_l R^{\rm max}, \qquad\quad\, l\in\{2, w\} \label{BRP1_Cons_5}\\
\sum_{k\in\mathcal{N}_h} x^{\prime k}_h R^{\prime k}_{l,h} = \beta_l R^{\rm max}, \qquad\quad l\in\{1, w\}. \label{BRP1_Cons_6}
\end{subnumcases}

In problem (P1), the objective function is to maximize the aggregated network utility. Since $\beta_1$, $\beta_2$, and $\beta_w$ are the only three slicing ratios, constraints (\ref{BRP1_Cons_1}) and (\ref{BRP1_Cons_2}) are considered in (P1). Constraints (\ref{BRP1_Cons_3}), (\ref{BRP1_Cons_4}), (\ref{BRP1_Cons_5}), and (\ref{BRP1_Cons_6}) indicate that spectrum resources allocated to AVs by a BS should be in this BS's available spectrum resources. According to problem (P1), each BS equally allocates its available spectrum resources to AVs associated to it (will be discussed in detail in the next section). However, the downlink transmission rate required by an AV depends on its application request. For a BS with a fixed amount of available spectrum, equally allocating spectrum to AVs associated to it and simultaneously guaranteeing their heterogeneous QoS requirements will reduce the number of AVs accessed to it. Thus, QoS constraints on each element in $\textbf{R}$ and $\textbf{R}^\prime$ are not considered in problem (P1) and the optimal $\{\beta_1^\star, \beta_2^\star, \beta_w^\star\}$ is regarded as the only output to determine the amount of spectrum resources reused by each BS.

\textbf{Spectrum allocating among AVs:} To accommodate situations with high density AVs, a linear network utility function is considered in spectrum allocating among AVs associated to the same BS. For given slicing ratios $\beta_1$, $\beta_2$, and $\beta_w$, and transmit power of each Wi-Fi AP, a network throughput maximization problem can be formulated as
\begin{equation}
\begin{split}
\label{P2}
\mathbf{P2}: \max_{\substack{\textbf{X}, \textbf{X}^\prime \\
                    \textbf{R}, \textbf{R}^\prime}} & \sum_{k\in \overline{\mathcal{M}}_1} \gamma^k_{1} + \sum_{W_g \in\mathcal{A}_1} \sum_{k\in \mathcal{N}_g} (x^k_{1} \gamma^k_{1} + x^{\prime k}_g \gamma^{\prime k}_g) \\
                     & + \sum_{k\in \overline{\mathcal{M}}_2} \gamma^k_{2} + \sum_{W_h \in\mathcal{A}_2} \sum_{k\in \mathcal{N}_h} (x^k_{2} \gamma^k_{2} + x^{\prime k}_h \gamma^{\prime k}_h) \\
\end{split}
\end{equation}
\addtocounter{equation}{-1}
\begin{subnumcases}
{\rm s.t.}
{\rm (\ref{BRP1_Cons_3})-(\ref{BRP1_Cons_6})} \\
R^k_1, R^k_2, R_{2,g}^{\prime k}, R_{w,g}^{\prime k}, R_{1,h}^{\prime k}, R_{w,h}^{\prime k} \geq 0  \label{BRP2_Cons_12} \\
x^k_1, x^k_2, x^{\prime k}_{g}, x^{\prime k}_{h} \in \{0,1\}, \qquad\qquad\ \ \,   k\in \mathcal{N}_i \label{BRP2_Cons_1} \\
x^k_1 + x^{\prime k}_g = 1, \qquad\qquad\qquad\quad k\in \cup_{W_g} \mathcal{N}_g \label{BRP2_Cons_2} \\
x^k_2 + x^{\prime k}_h = 1, \qquad\qquad\qquad\quad k\in \cup_{W_h} \mathcal{N}_h \label{BRP2_Cons_3} \\
\gamma^k_l\geq \gamma_{min}, \quad\ \, l\in\{1, 2\}, k\in \{\overline{\mathcal{M}}^s_1 \cup \overline{\mathcal{M}}_2^s \} \label{BRP2_Cons_4}\\
x^k_1[\gamma^k_1 - \gamma_{min}] \geq 0, \qquad\qquad\ \ k\in \cup_{W_g} \mathcal{N}^s_g  \label{BRP2_Cons_5}\\
x^k_2[\gamma^k_2 - \gamma_{min}] \geq 0,  \qquad\qquad\ \ k\in \cup_{W_h} \mathcal{N}^s_h  \label{BRP2_Cons_6}\\
x^{\prime k}_i [\gamma^{\prime k}_i - \gamma_{min}] \geq 0, \quad\,   k\in \cup_{W_i\in{\mathcal{A}_1\cup\mathcal{A}_2}} \mathcal{N}^s_i  \label{BRP2_Cons_7}\\
\gamma^k_l\geq \lambda_n L_n, \quad\,   l\in\{1, 2\}, k\in \{\overline{\mathcal{M}}^t_1 \cup \overline{\mathcal{M}}_2^t \} \label{BRP2_Cons_8}\\
x^k_1[\gamma^k_1 - \lambda_n L_n] \geq 0,  \qquad\qquad\,\,    k\in \cup_{W_g}  \mathcal{N}^t_g \label{BRP2_Cons_9}\\
x^k_2[\gamma^k_2 - \lambda_n L_n] \geq 0,  \qquad\qquad\,\,   k\in \cup_{W_h} \mathcal{N}^t_g \label{BRP2_Cons_10}\\
x^{\prime k}_i [\gamma^{\prime k}_i - \lambda_n L_n] \geq 0, \quad  k\in \cup_{W_i\in{\mathcal{A}_1\cup\mathcal{A}_2}}  \mathcal{N}^t_i \label{BRP2_Cons_11}
\end{subnumcases}
where $\textbf{X}=\{x^k_1, x^k_2\}$ and $\textbf{X}^\prime=\{x^{\prime k}_g, x^{\prime k}_h\}$ are the association matrices between eNBs and AVs, and between Wi-Fi APs and AVs, respectively; $L_n$ and $\lambda_n$ are the corresponding packet size and the arrival rate for delay-tolerant service requests; $\overline{\mathcal{M}}^s_j$/$\overline{M}^s_j$ (or $\overline{\mathcal{M}}^t_j$/$\overline{M}^t_j$) are set/number of AVs only within the coverage of eNB $S_j$ and request for delay-sensitive (or delay-tolerant) services; $\mathcal{N}^s_i$/$N^s_i$ (or $\mathcal{N}^t_i$/$N^t_i$) are set/number of AVs within the coverage of Wi-Fi AP $W_i$ and request for delay-sensitive (or delay-tolerant) services.

In problem (P2), the first four constraints are same with problem (P1) and used to demonstrate the required spectrum for each vehicle allocated from its associated BS with constraint (\ref{BRP2_Cons_12}) together. Constraints (\ref{BRP2_Cons_1})-(\ref{BRP2_Cons_3}) indicate that each vehicle is associated with either the eNB or the Wi-Fi AP closed to it. Constraints (\ref{BRP2_Cons_4})-(\ref{BRP2_Cons_7}) ensure the service rates from either an eNB and a Wi-Fi AP so that the delay requirement for a vehicle with delay-sensitive services can be guaranteed. For vehicles with delay-tolerant requests, constraints (\ref{BRP2_Cons_8})-(\ref{BRP2_Cons_11}) indicate that the service rate from eNBs or Wi-Fi APs should be not less than the periodic data traffic arrival rate at that eNB or Wi-Fi AP. Via solving problem (P2), the optimal association matrices for eNBs $\textbf{X}^\ast$ and for Wi-Fi APs $\textbf{X}^{\ast \prime}$, and local spectrum allocation matrices for eNBs $\textbf{R}^\ast$ and for Wi-Fi APs $\textbf{R}^{\ast \prime}$ can be obtained, which maximize the network throughput with guaranteed QoS for different AV applications.

\subsection{Transmit Power Control}
\label{subsec:Spe_Re_Slicing}

In addition to spectrum slicing and allocating among BSs and AVs, controlling the transmit power for Wi-Fi APs to adjust the inter-cell interference would further improve the spectrum utilization. Denote $\textbf{P}^{\prime}= \{P_i^{\prime}|W_i \in {\mathcal{A}_1\cup\mathcal{A}_2} \}$ as transmit power matrix of Wi-Fi APs. Equations (\ref{Spectrumeffi_S1}) and (\ref{Spectrumeffi_Sg2}) indicate that the received signal-to-interference-plus-noise (SINR) by vehicles from either an eNB or a Wi-Fi AP change with Wi-Fi APs' transmit powers, and therefore, impacting achievable transmission rates for the corresponding downlink. To obtain optimal transmit power control, similar to problem (P2), the linear utility function is considered in this part. For a given slicing ratio set $\{\beta_1, \beta_2, \beta_w\}$, BS-vehicle association pattern matrices $\textbf{X}$ and $\textbf{X}^\prime$, and local spectrum allocation matrices $\textbf{R}$ and $\textbf{R}^\prime$, the network throughput maximization problem focusing on transmit power control can be formulated as
\begin{equation}
\begin{split}
\label{P3}
\mathbf{P3}: \max_{\substack{\textbf{P}^{\prime}}} & \sum_{k\in \overline{\mathcal{M}}_1} \gamma^k_{1} + \sum_{W_g \in\mathcal{A}_1} \sum_{k\in \mathcal{N}_g} (x^k_{1} \gamma^k_{1} + x^{\prime k}_g \gamma^{\prime k}_g) \\
                     & + \sum_{k\in \overline{\mathcal{M}}_2} \gamma^k_{2} + \sum_{W_h \in\mathcal{A}_2} \sum_{k\in \mathcal{N}_h} (x^k_{2} \gamma^k_{2} + x^{\prime k}_h \gamma^{\prime k}_h) \\
\end{split}
\end{equation}
\addtocounter{equation}{-1}
\begin{subnumcases}
{\rm s.t.}
(\ref{BRP2_Cons_4})-(\ref{BRP2_Cons_11}), \label{BRP3_Cons_1}\\
P_i^{\prime} \in [0, P^{\rm max}], \qquad\  W_i\in \{\mathcal{A}_1\cup \mathcal{A}_2\} \label{BRP3_Cons_2}
\end{subnumcases}
where $P^{\rm max}$ is the maximum transmit power allowed by each Wi-Fi AP. In problem (P3), the first eight constraints in (\ref{BRP3_Cons_1}) are same with problem (P2) and used to ensure the QoS requirements for delay-sensitive and delay-tolerant services. Constraint (\ref{BRP3_Cons_2}) indicates that transmit power of each Wi-Fi AP is less than $P^{\rm max}$. Then the optimal transmit power for each Wi-Fi AP can be determined by solving problem (P3). From the above discussion, variables considered in problems (P1), (P2), and (P3) are coupled, thus the three problems should be solved jointly.

\section{Problem Analysis and Suboptimal Solution}
\label{sec:Pro_Analy}

Due to the binary variable matrices $\textbf{X}$ and $\textbf{X}^\prime$, problems (P2) and (P3) are combinatorial and difficult to solve. Thus, in this section, we first analyze each problem and then transform (P2) and (P3) to tractable forms before we jointly solving these three problems for the final optimal solutions.

\subsection{Problem Analysis}
\label{subsec:Pro-analy}

Let $\mathcal{N}_i^{\prime}$ be the set of AVs within and associated with Wi-Fi AP $W_i$, i.e., $\mathcal{N}_i^{\prime} = \{k\in \mathcal{N}_i | x^{\prime k}_i = 1\}$ for $W_i \in \{\mathcal{A}_1 \cup \mathcal{A}_2\}$, and $|\mathcal{N}_i^{\prime}|=N_i^{\prime}$. Then, the objective function of (P1) can be transformed into,
\begin{equation}
\begin{split}
\label{P1_Ob1}
& \sum_{k\in \{\mathcal{M}_1 \setminus (\cup_{W_g} \mathcal{N}_g^{\prime}) \}} {\rm log}(R^k_1 r^k_1)
+ \sum_{W_g} \sum_{k\in \mathcal{N}_g^{\prime}} {\rm log}(\gamma^{\prime k}_g) \\
& + \sum_{k\in \{\mathcal{M}_2 \setminus (\cup_{W_h} \mathcal{N}_h^{\prime}) \}} {\rm log}(R^k_2 r^k_2)
+ \sum_{W_h} \sum_{k\in \mathcal{N}_h^{\prime}} {\rm log}(\gamma^{\prime k}_h)
\end{split}
\end{equation}
where mathematical symbol, $\setminus$, describes the relative complement of one set with respect to another set. According to the constraints of (P1), the sets of spectrum allocating variables, $\{R^k_1\}$, $\{R^k_2\}$, $\{R_{2,g}^{\prime k}\}$, $\{R_{w,g}^{\prime k}\}$, $\{R_{1,h}^{\prime k}\}$, and $\{R_{w,h}^{\prime k}\}$, are independent with uncoupled constrains. Thus, similar to proposition 1 in \cite{Ye2018}, we can decompose problem (P1) into six subproblems and obtain the optimal fractions of spectrum allocated to AVs from the associated BSs as follows,
\begin{align}
\begin{split}
        \label{So.P1}
        R^{*}_1= R^{*k}_1=
        \frac{\beta_1 R^{\rm max}}{M_1 - \sum_{W_g}N_g^\prime} \\
        R^{*}_2= R^{*k}_2=
        \frac{\beta_2 R^{\rm max}}{M_2 - \sum_{W_h}N_h^\prime} \\
        R_{2,g}^{* \prime} = R_{2,g}^{* \prime k}=
        \frac{\beta_2 R^{\rm max}}{N_g^\prime} \\
        R_{w,g}^{* \prime} = R_{w,g}^{* \prime k}=
        \frac{\beta_w R^{\rm max}}{N_g^\prime} \\
        R_{1,h}^{* \prime} = R_{1,h}^{* \prime k}=
        \frac{\beta_1 R^{\rm max}}{N_h^\prime} \\
        R_{w,h}^{* \prime} = R_{w,h}^{* \prime k}=
        \frac{\beta_w R^{\rm max}}{N_h^\prime}. \\
\end{split}
\end{align}

Equation (\ref{So.P1}) indicates that each BS equally allocates spectrum to AVs associating to it. By replacing the spectrum allocating variables with Equation (\ref{So.P1}), problem (P1) can be transformed into
\begin{equation}
\begin{split}
\label{P1_1}
\mathbf{P1^{\prime}}: &  \max_{\substack{\beta_1, \beta_2, \beta_w}} \sum_{k\in \{\mathcal{M}_1 \setminus (\cup_{W_g} \mathcal{N}_g^{\prime}) \}} {\rm log}(\frac{\beta_1 R^{\rm max} r^k_1}{M_1 - \sum_{W_g}N_g^\prime})\\
& + \sum_{W_g} \sum_{k\in \mathcal{N}_g^{\prime}} {\rm log}( \frac{\beta_2 R^{\rm max} r_{2,g}^{\prime k} + \beta_w R^{\rm max} r_{w,g}^{\prime k}}{N_g^\prime}) \\
& + \sum_{k\in \{\mathcal{M}_2 \setminus (\cup_{W_h} \mathcal{N}_h^{\prime}) \}} {\rm log}( \frac{\beta_2 R^{\rm max} r^k_2}{M_2 - \sum_{W_h}N_h^\prime} ) \\
& + \sum_{W_h} \sum_{k\in \mathcal{N}_h^{\prime}} {\rm log}( \frac{\beta_1 R^{\rm max} r_{1,h}^{\prime k} + \beta_w R^{\rm max} r_{w,h}^{\prime k}}{N_h^\prime})
\end{split}
\end{equation}
\addtocounter{equation}{-1}
\begin{subnumcases}
{\rm s.t.}
{\rm (\ref{BRP1_Cons_1})-(\ref{BRP1_Cons_2})}  \label{BRP1_1_Cons_1}
\end{subnumcases}

Due to the binary variable matrices $\textbf{X}$ and $\textbf{X}^\prime$, using the brute force algorithm to solve problems (P2) and (P3) is with high complexity. To address this issue, we allow AVs within the overlapping coverage area of a Wi-Fi AP and an eNB to associate to one or both of the Wi-Fi AP and the eNB \cite{ye2013user}. Thus, binary matrices $\textbf{X}$ and $\textbf{X}^\prime$ are relaxed into real-valued matrices $\widetilde{\textbf{X}}$ and $\widetilde{\textbf{X}}^\prime$ with elements $\widetilde{x}_j^k\in [0,1]$ and $\widetilde{x}_i^{\prime k}\in [0,1]$, respectively. And then, we can transform problem (P2) into
\begin{equation}
\begin{split}
\label{P2_1}
\mathbf{P2^{\prime}}: \max_{\substack{\widetilde{\textbf{X}}, \widetilde{\textbf{X}}^\prime \\
                    \textbf{R}, \textbf{R}^\prime}} & \sum_{k\in \overline{\mathcal{M}}_1} \gamma^k_{1} + \sum_{W_g \in\mathcal{A}_1} \sum_{k\in \mathcal{N}_g} (\widetilde{x}^k_{1} \gamma^k_{1} + \widetilde{x}^{\prime k}_g \gamma^{\prime k}_g) \\
                     & + \sum_{k\in \overline{\mathcal{M}}_2} \gamma^k_{2} + \sum_{W_h \in\mathcal{A}_2} \sum_{k\in \mathcal{N}_h} (\widetilde{x}^k_{2} \gamma^k_{2} + \widetilde{x}^{\prime k}_h \gamma^{\prime k}_h) \\
\end{split}
\end{equation}
\addtocounter{equation}{-1}
\begin{subnumcases}
{\rm s.t.}
\sum_{k\in \overline{\mathcal{M}}_1} R^k_1 + \sum_{W_g} \sum_{k\in\mathcal{N}_g} \widetilde{x}^k_1 R^k_1 = \beta_1 R^{\rm max} \label{BRP2_1_Cons_13}\\
\sum_{k\in \overline{\mathcal{M}}_2} R^k_2 + \sum_{W_h} \sum_{k\in\mathcal{N}_h} \widetilde{x}^k_2 R^k_2 = \beta_2 R^{\rm max}
\label{BRP2_1_Cons_14}\\
\sum_{k\in\mathcal{N}_g} \widetilde{x}^{\prime k}_g R^{\prime k}_{l,g} = \beta_l R^{\rm max}, \qquad\quad\, l\in\{2, w\} \label{BRP2_1_Cons_15}\\
\sum_{k\in\mathcal{N}_h} \widetilde{x}^{\prime k}_h R^{\prime k}_{l,h} = \beta_l R^{\rm max}, \qquad\quad l\in\{1, w\} \label{BRP2_1_Cons_16} \\
R^k_1, R^k_2, R_{2,g}^{\prime k}, R_{w,g}^{\prime k}, R_{1,h}^{\prime k}, R_{w,h}^{\prime k} \geq 0   \label{BRP2_1_Cons_12} \\
\widetilde{x}^k_1, \widetilde{x}^k_2, \widetilde{x}^{\prime k}_{g}, \widetilde{x}^{\prime k}_{h} \in [0,1], \qquad\qquad\ \ \,   k\in \mathcal{N}_i \label{BRP2_1_Cons_1} \\
\widetilde{x}^k_1 + \widetilde{x}^{\prime k}_g = 1, \qquad\qquad\qquad\quad k\in \cup_{W_g} \mathcal{N}_g \label{BRP2_1_Cons_2} \\
\widetilde{x}^k_2 + \widetilde{x}^{\prime k}_h = 1, \qquad\qquad\qquad\quad k\in \cup_{W_h} \mathcal{N}_h \label{BRP2_1_Cons_3} \\
\gamma^k_l\geq \gamma_{min}, \quad\ \, l\in\{1, 2\}, k\in \{\overline{\mathcal{M}}^s_1 \cup \overline{\mathcal{M}}_2^s \} \label{BRP2_1_Cons_4}\\
\widetilde{x}^k_1[\gamma^k_1 - \gamma_{min}] \geq 0, \qquad\qquad\ \ k\in \cup_{W_g} \mathcal{N}^s_g  \label{BRP2_1_Cons_5}\\
\widetilde{x}^k_2[\gamma^k_2 - \gamma_{min}] \geq 0,  \qquad\qquad\ \ k\in \cup_{W_h} \mathcal{N}^s_h  \label{BRP2_1_Cons_6}\\
\widetilde{x}^{\prime k}_i [\gamma^{\prime k}_i - \gamma_{min}] \geq 0, \quad\,   k\in \cup_{W_i\in{\mathcal{A}_1\cup\mathcal{A}_2}} \mathcal{N}^s_i  \label{BRP2_1_Cons_7}\\
\gamma^k_l\geq \lambda_n L_n, \quad\,   l\in\{1, 2\}, k\in \{\overline{\mathcal{M}}^t_1 \cup \overline{\mathcal{M}}_2^t \} \label{BRP2_1_Cons_8}\\
\widetilde{x}^k_1[\gamma^k_1 - \lambda_n L_n] \geq 0,  \qquad\qquad\,\,    k\in \cup_{W_g}  \mathcal{N}^t_g \label{BRP2_1_Cons_9}\\
\widetilde{x}^k_2[\gamma^k_2 - \lambda_n L_n] \geq 0,  \qquad\qquad\,\,   k\in \cup_{W_h} \mathcal{N}^t_g \label{BRP2_1_Cons_10}\\
\widetilde{x}^{\prime k}_i [\gamma^{\prime k}_i - \lambda_n L_n] \geq 0, \quad  k\in \cup_{W_i\in{\mathcal{A}_1\cup\mathcal{A}_2}}  \mathcal{N}^t_i. \label{BRP2_1_Cons_11}
\end{subnumcases}

To analyze the concavity property of problems (P1$^\prime$) and (P2$^\prime$), three definitions about concave functions \cite{boyd2004convex, gorski2007biconvex} and two concavity-preserving operations \cite{boyd2004convex} are introduced in Appendix \ref{Appendix:Def1}. The following propositions, proved in Appendix \ref{Appendix:Pro2} and Appendix \ref{Appendix:Pro3}, summarize the concavity property of problems (P1$^\prime$) and (P2$^\prime$), respectively,
\begin{myProp}
The objective function of problem (P1$^\prime$) is a concave function on the three optimal variables $\beta_1$, $\beta_2$, and $\beta_w$, and problem (P1$^\prime$) is a concave optimization problem.
\end{myProp}

\begin{myProp}
The objective function of problem (P2$^\prime$) is a biconcave function on variable set $\{\widetilde{\textbf{X}}, \widetilde{\textbf{X}}^\prime\}\times \{\textbf{R}, \textbf{R}^\prime\}$, and problem (P2$^\prime$) is a biconcave optimization problem.
\end{myProp}

Even though the integer-value variables in problem (P3) can be relaxed to real-value ones by replacing constraint (\ref{BRP3_Cons_1}) by (\ref{BRP2_1_Cons_4})-(\ref{BRP2_1_Cons_11}), the non-concave or non-biconcave relations between the objective function and decision variable of problem (P3) makes it difficult to solve directly. Thus, we use the first-order Taylor series approximation, and introduce two new variable matrices, $\textbf{C} = \{C^k_1, C^k_2\}$ and $\textbf{C}^\prime = \{C^{\prime k}_{2,g}, C^{\prime k}_{w,g}, C^{\prime k}_{1,h}, C^{\prime k}_{w,h}\}$ with elements that are linear-fractional function of $P_i^{\prime}$, to replace the received SINR on AVs within each BS's coverage area. Then, the downlink spectrum efficiency on an AV associated to a BS can be re-expressed as a concave function of $C$. For example, using $C^k_1$ to replace the SINR received on AV $k$ associated to eNB $S_1$, we can rewritten equation (\ref{Spectrumeffi_S1}) as
\begin{align}
\begin{split}
        \label{Spectrumeffi_S1_new}
        r^k_1 = {\rm log}_2 (1+ C^k_1).
\end{split}
\end{align}
Therefore, problem (P3) can be transformed into
\begin{equation}
\begin{split}
\label{P3_1}
\mathbf{P3^\prime}: & \max_{\substack{\textbf{P}^{\prime}, \textbf{C}, \textbf{C}^\prime}} \sum_{k\in \overline{\mathcal{M}}_1} R^k_1 {\rm log}_2 (1+ C^k_1) + \sum_{k\in \overline{\mathcal{M}}_2} R^k_2{\rm log}_2 (1\\
&+ C^k_2) + \sum_{W_g \in\mathcal{A}_1} \sum_{k\in \mathcal{N}_g} (x^k_{1} R^k_1 {\rm log}_2 (1+ C^k_1) + x^{\prime k}_g \\
& (R^{\prime k}_{2,g}{\rm log}_2 (1+ C^{\prime k}_{2,g}) + R^{\prime k}_{w,g} {\rm log}_2 (1+ C^{\prime k}_{w,g}))) \\
& + \sum_{W_h \in\mathcal{A}_2} \sum_{k\in \mathcal{N}_h} (x^k_{2} R^k_2 {\rm log}_2 (1+ C^k_2) + x^{\prime k}_h (R^{\prime k}_{1,h} \\
& {\rm log}_2 (1+ C^{\prime k}_{1,h}) + R^{\prime k}_{w,h} {\rm log}_2 (1+ C^{\prime k}_{w,h}))) \\
\end{split}
\end{equation}
\addtocounter{equation}{-1}
\begin{subnumcases}
{\rm s.t.}
{\rm (\ref{BRP2_1_Cons_4})-(\ref{BRP2_1_Cons_11})} \label{BRP3_1_Cons_1}\\
P_i^{\prime} \in [0, P^{\rm max}], \qquad\  W_i\in \{\mathcal{A}_1\cup \mathcal{A}_2\} \label{BRP3_1_Cons_2}\\
C^k_1 \leq \xi^k_1 \label{BRP3_1_Cons_3}\\
C^k_2 \leq \xi^k_2 \label{BRP3_1_Cons_4}\\
C^{\prime k}_{2,g} \leq \xi^{\prime k}_{2,g} \label{BRP3_1_Cons_5}\\
C^{\prime k}_{w,g} \leq \xi^{\prime k}_{w,g} \label{BRP3_1_Cons_6}\\
C^{\prime k}_{1,h} \leq \xi^{\prime k}_{1,h} \label{BRP3_1_Cons_7}\\
C^{\prime k}_{w,h} \leq \xi^{\prime k}_{w,h} \label{BRP3_1_Cons_8}
\end{subnumcases}
where $\xi^k$ (or $\xi^{\prime k}$) are the received SINRs on AV $k$ from its associated eNB (or Wi-Fi AP). The six additional constraints (\ref{BRP3_1_Cons_3})-(\ref{BRP3_1_Cons_8}) are biaffine on $\{\textbf{P}^{\prime}\}\times \{\textbf{C}, \textbf{C}^\prime \}$ and are considered in problem (P3$^\prime$) to ensure the equivalent with problems (P3).

\subsection{Algorithms Design}
\label{subsec:Algo_Design}

To jointly solve the three problems (P1$^\prime$), (P2$^\prime$), and (P3$^\prime$), we first design an alternate algorithm for (P3$^\prime$) and then an \emph{alternate concave search} (ACS) algorithm is applied to jointly solve these three problems. For simplicity, the objective functions for the three problems are denoted by $\mathcal{U}_{(P1^\prime)}$, $\mathcal{U}_{(P2^\prime)}$, and $\mathcal{U}_{(P3^\prime)}$, respectively.

The objective function of problem (P3$^\prime$), $\mathcal{U}_{(P3^\prime)}$, is concave on $\{\textbf{C}, \textbf{C}^\prime \}$, while constraints (\ref{BRP3_1_Cons_3})-(\ref{BRP3_1_Cons_8}) are biaffine on $\{\textbf{P}^{\prime}\}\times \{\textbf{C}, \textbf{C}^\prime \}$. Through maximizing $\mathcal{U}_{(P3^\prime)}$, optimal $\{\textbf{C}, \textbf{C}^\prime \}$ can be obtained for given $\textbf{P}^{\prime}$ with constraints (\ref{BRP3_1_Cons_3})-(\ref{BRP3_1_Cons_8}). Moreover, through maximizing $0$ with constraints (\ref{BRP3_1_Cons_1})-(\ref{BRP3_1_Cons_8}), the feasible set of $\textbf{P}^{\prime}$ can be obtained. Thus, we first separate problem (P3$^\prime$) into two subproblems as follows
\begin{equation}
\begin{split}
\label{P3_1}\nonumber
\mathbf{P3^\prime.SP1}: & \max_{\substack{\textbf{C}, \textbf{C}^\prime}} \,\ \mathcal{U}_{(P3^\prime)} \\
 & {\rm s.t.} \,\ (\ref{BRP3_1_Cons_3})-(\ref{BRP3_1_Cons_8})
\end{split}
\end{equation}
and
\begin{equation}
\begin{split}
\label{P3_1}\nonumber
\mathbf{P3^\prime.SP2}: & \max_{\substack{\textbf{P}^{\prime}}} \,\ 0 \\
& {\rm s.t.} \,\ (\ref{BRP3_1_Cons_1})-(\ref{BRP3_1_Cons_8}).
\end{split}
\end{equation}
It is obvious that there must be a solution to subproblem (P3$^\prime$.SP1). Moreover, since subproblem (P3$^\prime$.SP2) is a feasibility problem and the initial value of \textbf{P}$^{\prime}$ is always the solution for (P3$^\prime$.SP2). Thus, problem (P3$^\prime$) converges and can be solved by iteratively solving subproblems (P3$^\prime$.SP1) and (P3$^\prime$.SP2).

To jointly solve (P1$^\prime$), (P2$^\prime$), and (P3$^\prime$) and obtain the final optimal decision variables, the ACS algorithm is summarized in Algorithm 1. $\{\widetilde{\textbf{X}}^{(t)}, \widetilde{\textbf{X}}^{(t)\prime}\}$ and \textbf{P}$^{(t)\prime}$ are the values of $\{\widetilde{\textbf{X}}, \widetilde{\textbf{X}}^\prime\}$ and \textbf{P}$^{\prime}$ at the beginning of the $t$th iteration, and $\mathcal{U}^{(t)}_{(P2^\prime)}$ is the maximum objective function value of problem (P2$^\prime$) with optimal decision variables $\{\beta_1^{(t)}, \beta_2^{(t)}, \beta_w^{(t)}\}$, $\{\widetilde{\textbf{R}}^{(t)}, \widetilde{\textbf{R}}^{(t)\prime}\}$, $\{\widetilde{\textbf{X}}^{(t)}, \widetilde{\textbf{X}}^{(t)\prime}\}$, and \textbf{P}$^{(t)\prime}$. To enhance the convergence speed of Algorithm 1, the output at the $(t-1)$th iteration is regarded as a feedback to the input at the $t$th iteration \cite{guo2018fast}, such as, the $t$th input $\textbf{P}^{(t)\prime}$ is defined as
\begin{equation}
\begin{split}
\label{input_P}
\textbf{P}^{(t)\prime}=\textbf{P}^{(t-1)\prime} + \theta(\textbf{P}^{\dag\prime} - \textbf{P}^{(t-1)\prime})
\end{split}
\end{equation}
where, $\theta$ is the feedback coefficient. Moreover, considering that a lager $\theta$ may result in missing optimal output at each iteration while a small $\theta$ reduces the convergence speed, two coefficients $\theta_1$ and $\theta_2$ are considered in Algorithm 1.

According to the analysis of each problem in subsection \ref{subsec:Pro-analy}, Algorithm 1 converges since:
\begin{enumerate}[(i)]
\item The output of problems (P1$^\prime$) and (P2$^\prime$), $\{\beta_1, \beta_2, \beta_w\}$, $\{\widetilde{\textbf{X}}, \widetilde{\textbf{X}}^\prime\}$, and $\{\widetilde{\textbf{R}}, \widetilde{\textbf{R}}^\prime\}$, are closed sets;
\item Both (P1$^\prime$) and (P2$^\prime$) are concave/biconcave optimization problems such that the optimal solution for each problem at the end of the $k$th iteration is unique when the input of the algorithm is the optimal results obtained from the $(k-1)$th iteration;
\item Problem (P3$^\prime$) is always solvable.
\end{enumerate}

\begin{algorithm}
\SetAlgorithmName{Algorithm}{List of Models}
\DontPrintSemicolon
\KwIn{Input parameters for (P1$^\prime$), (P2$^\prime$), and (P3$^\prime$); initial values for $\{\widetilde{\textbf{X}}, \widetilde{\textbf{X}}^\prime\}$ and \textbf{P}$^{\prime}$; stopping criterion $\kappa_1$; feedback coefficient updating criterion $\kappa_2$ ($\kappa_2 > \kappa_1$); feedback coefficients $\theta_1$ and $\theta_2$; maximum iterations $\widehat{N}$.}
\KwOut{Optimal spectrum slicing ratios, $\{\beta_1^\ast, \beta_2^\ast, \beta_w^\ast\}$; Optimal local spectrum allocation matrix, $\{\textbf{R}^\ast, \textbf{R}^{\ast \prime}\}$; Optimal BS-vehicle association patterns, $\{\widetilde{\textbf{X}}^\ast, \widetilde{\textbf{X}}^{\ast\prime}\}$; Optimal transmit powers for APs, \textbf{P}$^{\ast \prime}$; Optimal SINR matrices $\{\textbf{C}^\ast, \textbf{C}^{\ast \prime}\}$.}
\tcc{Initialization phase}
\For{the first iteration, $k=0$}{
set initial values for $\{\widetilde{\textbf{X}}, \widetilde{\textbf{X}}^\prime\}$ and \textbf{P}$^{\prime}$, denoted by $\{\widetilde{\textbf{X}}^{(0)}, \widetilde{\textbf{X}}^{(0)\prime}\}$ and \textbf{P}$^{(0)\prime}$, respectively; set $\mathcal{U}^{(0)}_{(P2^\prime)}$ to $0$.
}
\tcc{Solving iteratively phase}
\Repeat{$\parallel \mathcal{U}^{(t)}_{(P2^\prime)} - \mathcal{U}^{(t-1)}_{(P2^\prime)}\parallel \leq \kappa_1$ or $k\geq \widehat{N}$}{
 \ForEach{$k\leq \widehat{N}$}{
 Step1: $\{\beta_1^\dag, \beta_2^\dag, \beta_w^\dag\}$ $\leftarrow$ solving (P1$^\prime$) given $\{\widetilde{\textbf{X}}^{(t)}, \widetilde{\textbf{X}}^{(t)\prime}\}$ and \textbf{P}$^{(t)\prime}$; \\
 Step2: $\{\textbf{R}^{\dag}, \textbf{R}^{\dag \prime}\}$ $\leftarrow$ solving (P2$^\prime$) given $\{\beta_1^\dag, \beta_2^\dag, \beta_w^\dag\}$, $\{\widetilde{\textbf{X}}^{(t)}, \widetilde{\textbf{X}}^{(t)\prime}\}$, and \textbf{P}$^{(t)\prime}$; \\
 Step3: $\{\widetilde{\textbf{X}}^{\dag}, \widetilde{\textbf{X}}^{\dag \prime}\}$ $\leftarrow$ solving (P2$^\prime$) given $\{\beta_1^\dag, \beta_2^\dag, \beta_w^\dag\}$, $\{\textbf{R}^{\dag}, \textbf{R}^{\dag \prime}\}$, and \textbf{P}$^{(t)\prime}$; \\
 Step4: $\{\textbf{C}^{(t+1)}, \textbf{C}^{(t+1) \prime}\}$, \textbf{P}$^{(t+1) \prime}$ $\leftarrow$ solving (P3$^\prime$) by iteratively solving (P3$^\prime$.SP1) and (P3$^\prime$.SP2) given $\{\beta_1^\dag, \beta_2^\dag, \beta_w^\dag\}$, $\{\textbf{R}^{\dag}, \textbf{R}^{\dag \prime}\}$, and $\{\widetilde{\textbf{X}}^{\dag}, \widetilde{\textbf{X}}^{\dag \prime}\}$; \\
     \uIf{No feasible solutions for (P1$^\prime$), (P2$^\prime$), or (P3$^\prime$)}{
     Go to initialization phase and reset the initial values for related parameters until no feasible solutions found;
     Stop and no optimal solutions under current network setting;
		    	}
\ElseIf{$\parallel \mathcal{U}^{(t)}_{(P2^\prime)} - \mathcal{U}^{(t-1)}_{(P2^\prime)}\parallel \leq \kappa_2$}{
$\{\beta_1^{(t+1)}, \beta_2^{(t+1)}, \beta_w^{(t+1)}\}$ $\leftarrow$ $\{\beta_1^{(t)}, \beta_2^{(t)}, \beta_w^{(t)}\} + \theta_2*(\{\beta_1^\dag, \beta_2^\dag, \beta_w^\dag\} - \{\beta_1^{(t)}, \beta_2^{(t)}, \beta_w^{(t)}\})$; \\
$\{\textbf{R}^{(t+1)}, \textbf{R}^{(t+1) \prime}\}$ $\leftarrow$ $\{\textbf{R}^{(t)}, \textbf{R}^{(t) \prime}\} + \theta_2*( \{\textbf{R}^{\dag}, \textbf{R}^{\dag \prime}\} - \{\textbf{R}^{(t)}, \textbf{R}^{(t) \prime}\})$;\\
$\{\textbf{X}^{(t+1)}, \textbf{X}^{(t+1) \prime}\}$ $\leftarrow$ $\{\textbf{X}^{(t)}, \textbf{X}^{(t) \prime}\} + \theta_2*( \{\textbf{X}^{\dag}, \textbf{X}^{\dag \prime}\} - \{\textbf{X}^{(t)}, \textbf{X}^{(t) \prime}\})$.}

\Else{$\{\beta_1^{(t+1)}, \beta_2^{(t+1)}, \beta_w^{(t+1)}\}$ $\leftarrow$ $\{\beta_1^{(t)}, \beta_2^{(t)}, \beta_w^{(t)}\} + \theta_1*(\{\beta_1^\dag, \beta_2^\dag, \beta_w^\dag\} - \{\beta_1^{(t)}, \beta_2^{(t)}, \beta_w^{(t)}\})$; \\
$\{\textbf{R}^{(t+1)}, \textbf{R}^{(t+1) \prime}\}$ $\leftarrow$ $\{\textbf{R}^{(t)}, \textbf{R}^{(t) \prime}\} + \theta_1*( \{\textbf{R}^{\dag}, \textbf{R}^{\dag \prime}\} - \{\textbf{R}^{(t)}, \textbf{R}^{(t) \prime}\})$;\\
$\{\textbf{X}^{(t+1)}, \textbf{X}^{(t+1) \prime}\}$ $\leftarrow$ $\{\textbf{X}^{(t)}, \textbf{X}^{(t) \prime}\} + \theta_1*( \{\textbf{X}^{\dag}, \textbf{X}^{\dag \prime}\} - \{\textbf{X}^{(t)}, \textbf{X}^{(t) \prime}\})$.}

 Obtain $\mathcal{U}^{(t+1)}_{(P2^\prime)}$ at the end of $k$th iteration with $\{\beta_1^{(t+1)}, \beta_2^{(t+1)}, \beta_w^{(t+1)}\}$, $\{\widetilde{\textbf{R}}^{(t+1)}, \widetilde{\textbf{R}}^{(t+1)\prime}\}$, $\{\widetilde{\textbf{X}}^{(t+1)}, \widetilde{\textbf{X}}^{(t+1)\prime}\}$, and \textbf{P}$^{(t+1)\prime}$;
  } $k$ $\leftarrow$ $k+1$;
  }
\caption{The ACS algorithm for jointly solving (P1$^\prime$), (P2$^\prime$), and (P3$^\prime$) \label{Model:Algorithm2}}
\end{algorithm}

%

\section{Simulation Results}
\label{sec:simmu}

To show the effectiveness of our proposed spectrum resource management framework, extensive simulation is carried out. We compare the proposed spectrum resource management scheme with two existing resource slicing schemes, i.e., the maximization-utility (max-utility) based resource slicing scheme proposed in \cite{Ye2018}, and the maximization-SINR (max-SINR) based resource slicing scheme proposed in \cite{ye2013user}. 
The BS-user association patterns and spectrum slicing ratios are optimized with objective of maximizing the aggregated network utility in max-utility scheme while users choose to associate with the BS providing higher SINR and only spectrum slicing ratios are optimized in max-SINR scheme.

We consider two eNBs (eNB $S_1 \in \mathcal{B}_1$ and eNB $S_2 \in \mathcal{B}_2$) and four Wi-Fi APs (AP $1$ and AP $2$ in $\mathcal{A}_1$, and AP $3$ and AP $4$ in $\mathcal{A}_2$) are utilized for AV applications. Transmit power is fixed at $10\,$watts (i.e., $40\,$dBm) for each eNB with a maximum communication range of $600\,$m. Since no transmit power control for both of max-utility and max-SINR schemes, transmit powers of APs are set as $1\,$watt with communication range of $200\,$m, the same as in \cite{Ye2018}. In our simulation, the minimum inter-vehicle distance is $5\,$m, and the AV density over one lane, i.e., the number of AVs on one lane per meter, varies within range of $[0.04, 0.20]\,$AV/m. The downlink channel gains for eNBs and Wi-Fi APs are described as $L_e(d)= -30-35 {\rm log}_{_{10}}(d)$ and $L_w(d)= -40-35 {\rm log}_{_{10}}(d)$ \cite{Ye2018}, respectively, where $d$ is the BS-vehicle distance. We take platooning/convoying as an example to set the delay bound for delay-sensitive applications, i.e., $10\,$ms \cite{peng2018vehicular, lema2017business}, and downloading HD map is considered as an example for delay-tolerant applications \cite{yuan2018toward}. Other important parameters in our simulation are listed in Table \ref{table:parameters}.

\begin{table}[htbp]
\setlength{\belowcaptionskip}{5pt}
 \caption{\label{table:parameters} Parameters values}
 \centering
 \begin{tabular}{p{5.3cm}||p{2.2cm}}
  \toprule
  \textbf{Parameter} & \textbf{Value}\\
  \midrule
Maximum transmit power allowed by APs & $2.5\,$watts \\
Background noise power & $-104\,$dBm \\
HD map packet arrival rate & $20\,$packet/s \\
HD map packet size & $9000\,$bits \\
Safety-sensitive packet arrival rate & $4\,$packet/s \\
Safety-sensitive packet size & $1048\,$bits \\
Safety-sensitive packet delay bound & $10\,$ms \\
Safety-sensitive request generating probability & $0.1-0.9$ \\
Delay bound violation probability & $10^{-3}$ \\
$\theta_1$/$\theta_2$ & 0.001/0.1 \\
$\kappa_1$/$\kappa_2$ & 0.01/20 \\
  \bottomrule
 \end{tabular}
\end{table}

\begin{figure}[htbp]
\centering
\subfigure[$p=0.2$]{
\label{fig:CapVSWvP2}
\includegraphics[width=0.38\textwidth]{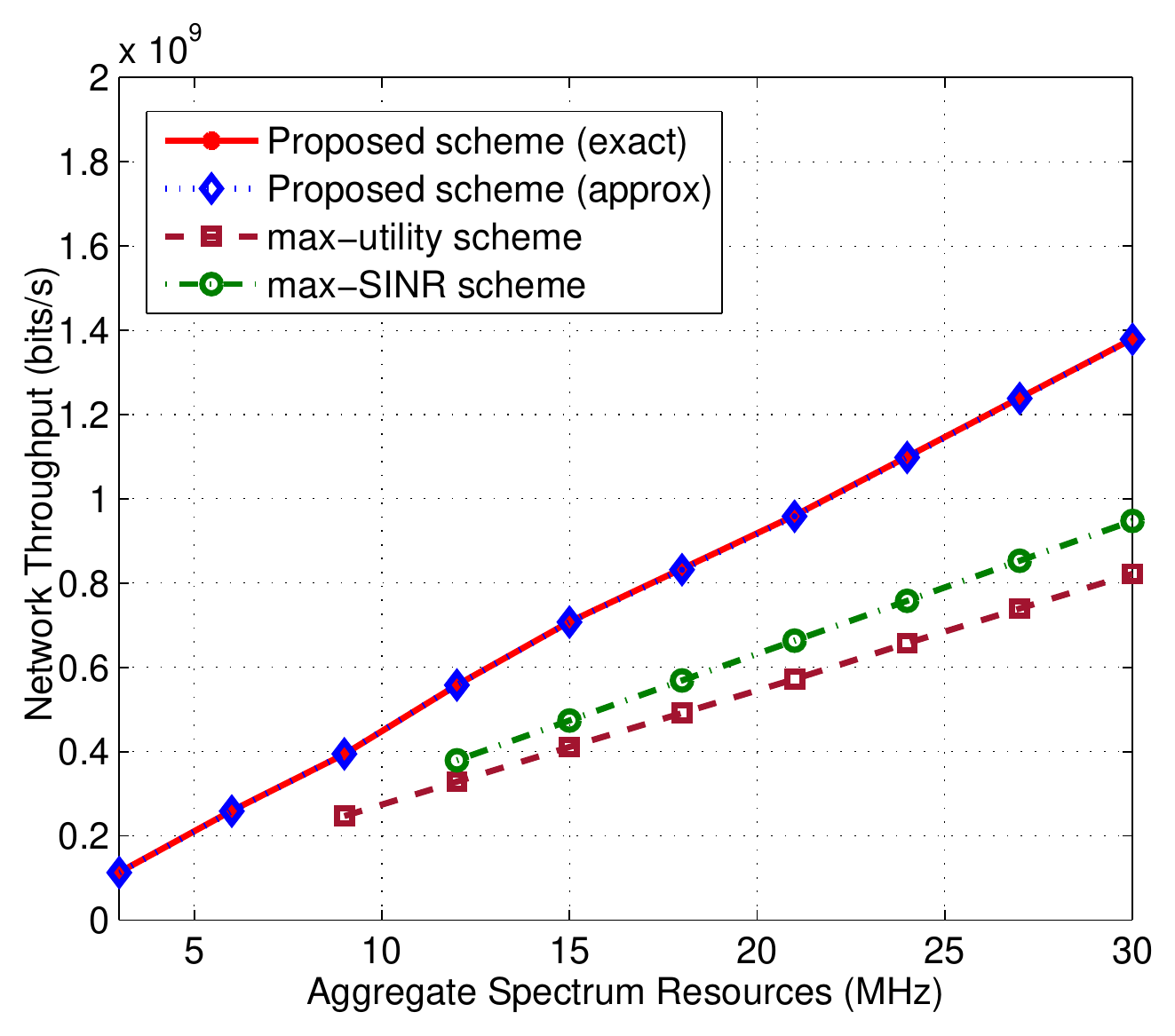}}
\subfigure[$p=0.8$]{
\label{fig:CapVSWvP8}
\includegraphics[width=0.38\textwidth]{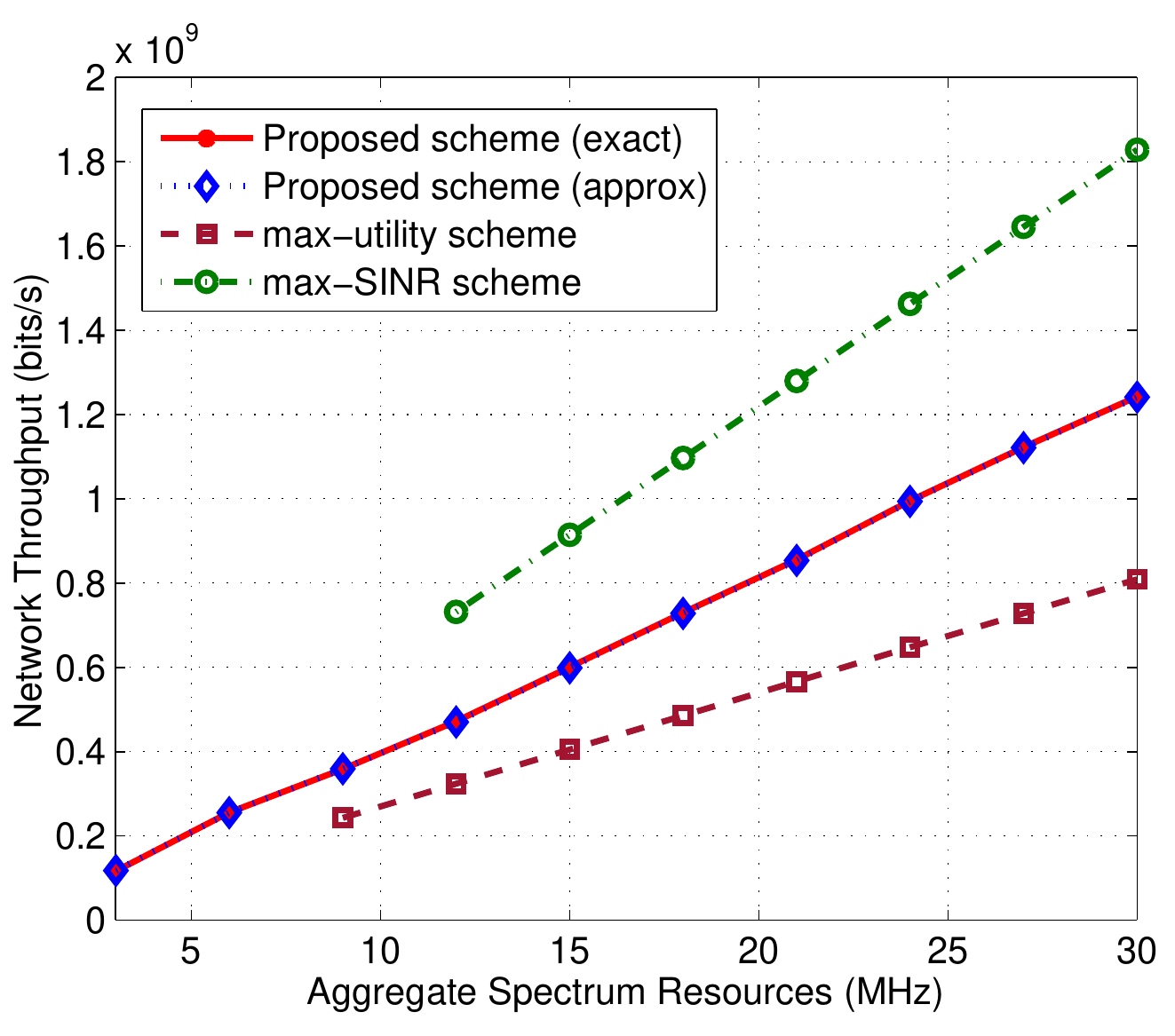}}
\caption{Comparison of network throughput vs. aggregate spectrum resources under the same AV distribution with AV density $0.05$AV/m.}
\label{fig:CapVSWv}
\end{figure}

\begin{figure}[htbp]
\centering
\includegraphics[height=2.5 in]{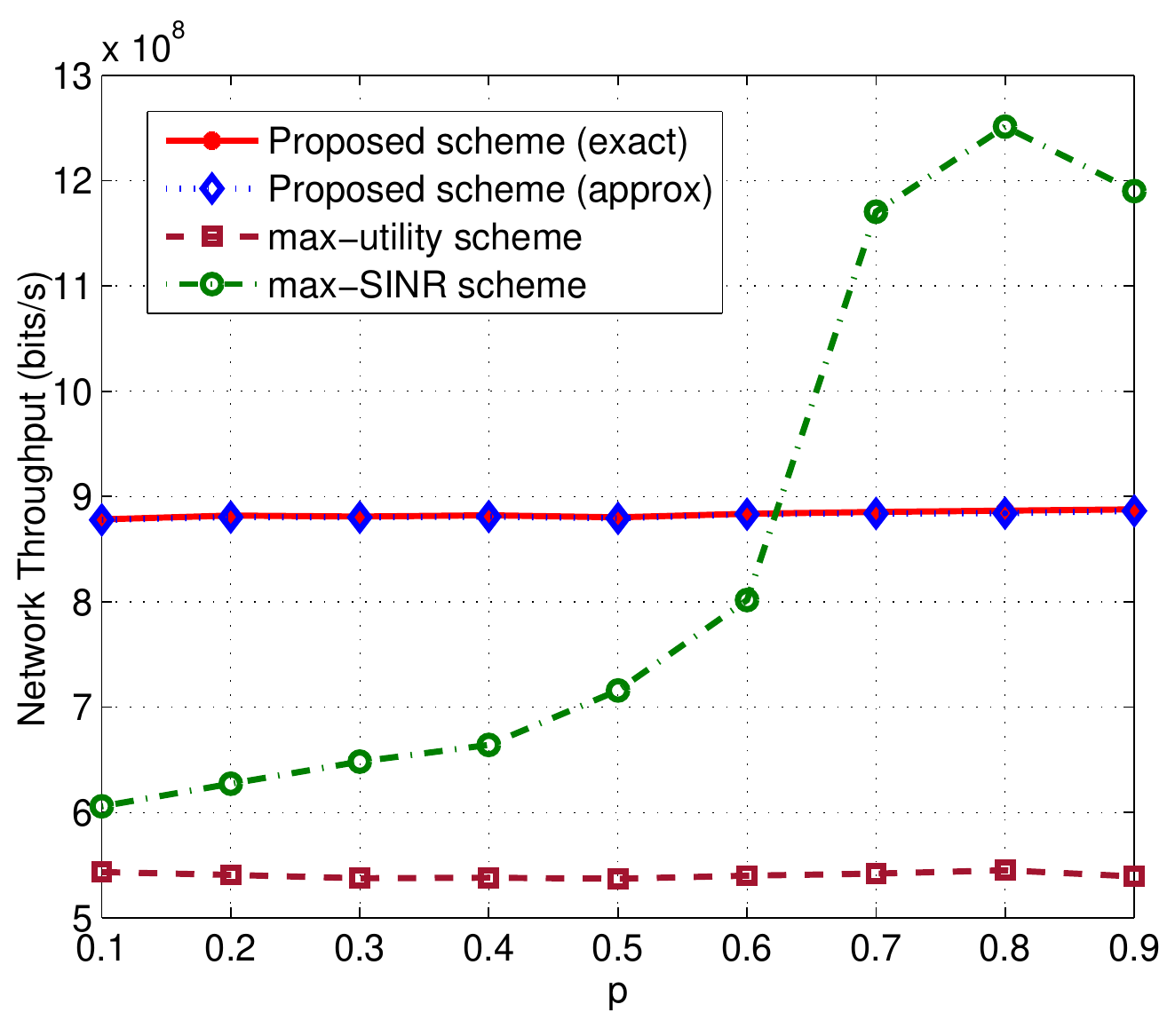}
\caption{Average network throughput vs. $p$ (AV density is $0.05$AV/m).}
\label{fig:CapVSp}
\end{figure}

\begin{figure}[htbp]
\centering
\subfigure[$p=0.2$]{
\label{fig:CapVSDensp2}
\includegraphics[width=0.38\textwidth]{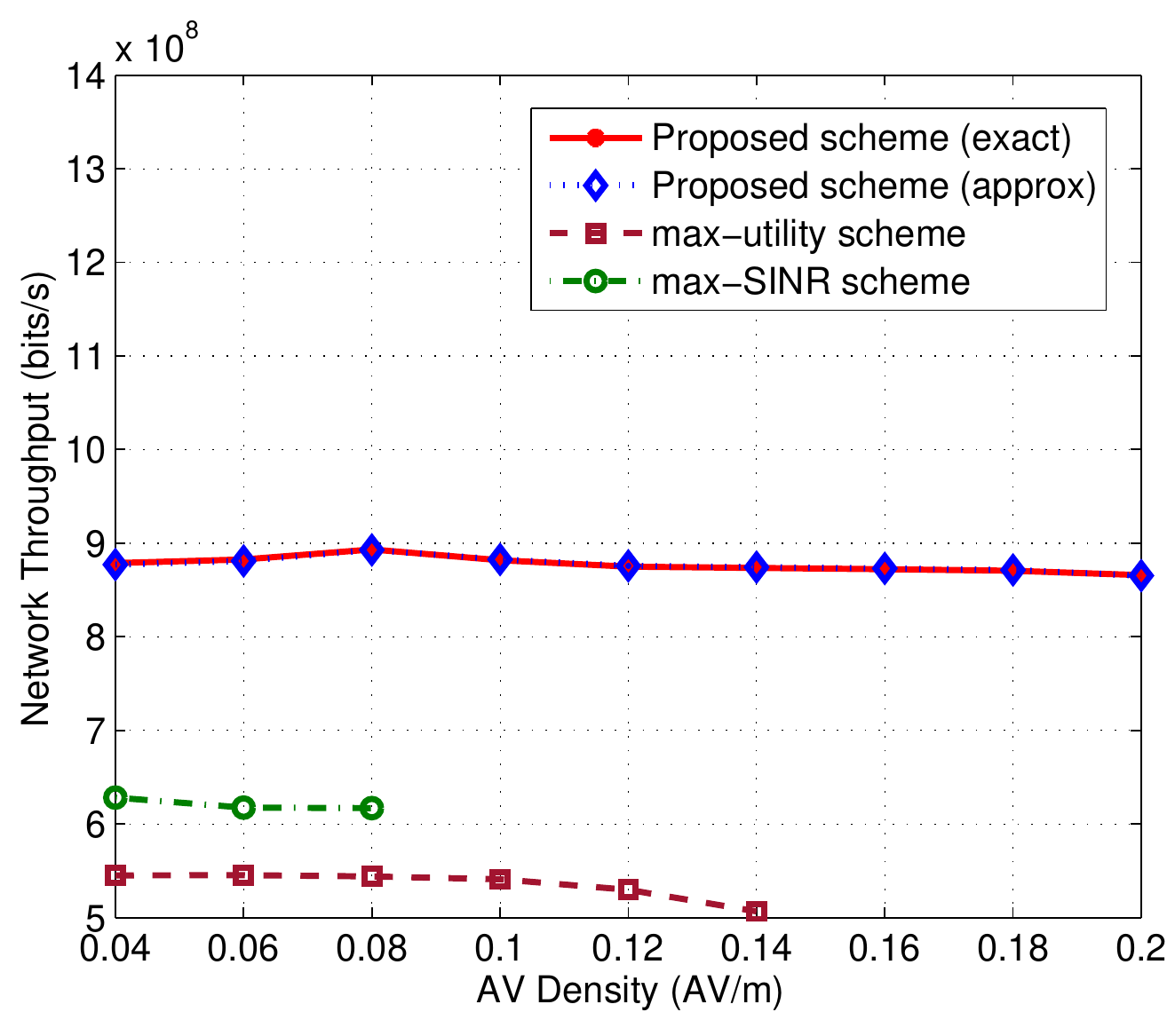}}
\subfigure[$p=0.8$]{
\label{fig:CapVSDensp8}
\includegraphics[width=0.38\textwidth]{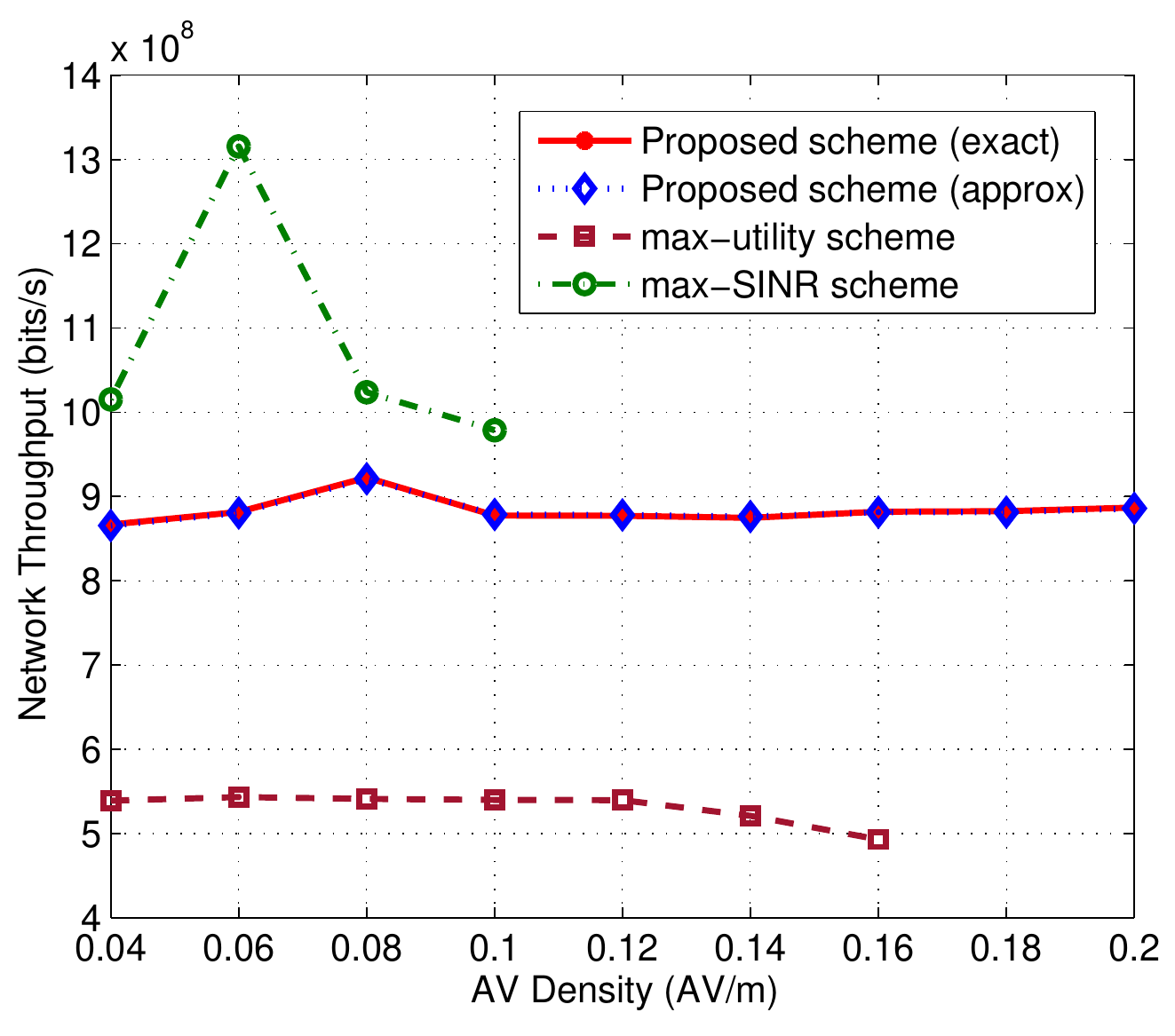}}
\caption{Average network throughput vs. AV density.}
\label{fig:CapVSDens}
\end{figure}

We use network throughput that is, the summation of achievable transmission rate by each individual AV from BSs, to measure performances of different spectrum resource management schemes. Considering the scarcity of spectrum resources, the different vehicular applications, and the high dynamic of vehicular networks, we evaluate the performance of the proposed scheme and compare with max-utility and max-SINR schemes under different amounts of aggregate spectrum resource ($W_v$), probabilities of generating a delay-sensitive request by AVs ($p$), and AV densities in Fig. \ref{fig:CapVSWv} to Fig. \ref{fig:CapVSDens}.

Fig. \ref{fig:CapVSWv} demonstrates the network throughputs achieved by the three schemes with respect to different amounts of aggregate spectrum resources, $W_v$, where AV density is $0.05\,$AV/m and $p=0.2$ and $0.8$, respectively. 
With the increasing of $W_v$, transmission rate for each AV is increased due to the increasing of the amount of allocated spectrum resources. 
From Fig. \ref{fig:CapVSWv}, the minimum requirement for spectrum resources by the proposed scheme to support the downlink transmissions 
is $3\,$MHz while at least $9\,$MHz and $12\,$MHz spectrum are required by the max-utility scheme and the max-SINR scheme, respectively. Moreover, under different $W_v$, the network throughput achieved by the proposed scheme is on average over $70\%$ and over $50\%$ higher than that of the max-utility scheme for $p=0.2$ and $0.8$, respectively, and over $45\%$ higher on average than that of the max-SINR scheme for $p=0.2$. From Fig. \ref{fig:CapVSWvP2}, 
with the increase of $W_v$, network throughput achieved by the proposed scheme increases more rapidly than the max-utility scheme.

Network throughputs of the three schemes under different $p$ are evaluated in Fig. \ref{fig:CapVSp}. The effect of $p$ on network throughput is mainly caused by the difference between the QoS requirements for delay-sensitive and delay-tolerant applications. According to Equation (\ref{minrate}) and the parameter setting in Table \ref{table:parameters}, the transmission rate required by a delay-tolerant request is $180.00\,$kbits/s, which is higher than that for a delay-sensitive request, $140.37\,$kbits/s. A large $p$ indicates a low total transmission rate required by all AVs to satisfy their applications' QoS requirements, therefore more remaining spectrum resources can be allocated to AVs with higher received SINRs in the proposed scheme. Thus, under the scenarios with the same AV density, $0.05\,$AV/m, network throughputs of the three schemes increase with $p$. For the max-SINR scheme, AVs associate the BS providing higher SINR and each BS equally allocates spectrum to AVs. To guarantee the QoS requirements for AVs, the amount of resource allocated to AVs from the same BS fluctuates with the distribution of BS-vehicle SINR and $p$, resulting in drastic impact on the achieved network throughput. Moreover, from Fig. \ref{fig:CapVSp}, the proposed scheme outperforms the max-SINR scheme when $p$ is small and can achieve higher network throughput than the max-utility scheme for the scenario with different $p$.

Fig. \ref{fig:CapVSDens} shows the network throughputs of the three schemes under different AV densities with $p=0.2$ and $0.8$, respectively, and $20\,$MHz aggregate spectrum resources. From the figure, the proposed scheme is more robust to AV density changing than the other two. 
For both the max-SINR and the max-utility schemes, 
only scenarios with relatively small AV densities can be accommodated due to equal spectrum allocation among AVs and unbalance between the downlink data traffic and the available aggregate spectrum resources. 
Furthermore, 
the proposed scheme has over $50\%$ increase in the achieved network throughput than the max-utility scheme with $p=0.2$ and $0.8$ and has over $40\%$ increase than the max-SINR scheme when $p=0.2$.

Fig. \ref{fig:CapVSDens} also indicates the effect of AV density on the achieved network throughputs by the three schemes. 
In general, the network throughputs achieved by the three schemes overall decrease with AV density. 
To increase the network throughput, the proposed scheme and the max-SINR scheme prefer to slicing high spectrum ratio to the BSs providing higher SINRs to its associated AVs once enough spectrum is allocated to each AV to guarantee the QoS requirements for their applications. 
However, the amount of spectrum resource needed to satisfy AV application's QoS requirements increases with the AV density for the three schemes, thus less spectrum resources can be used for increasing network throughput, resulting in decreasing in network throughput.


\begin{table*}[htbp]
  \centering
  \fontsize{7.5}{9}\selectfont
  \begin{threeparttable}
  \caption{Optimal transmit powers and number of iterations for the three schemes ($p=0.8$).}
  \label{tab:performance_comparison}
    \begin{tabular}{cccccccc}
    \toprule
    \multirow{2}{*}{AV Density (AV/m)}&
    \multicolumn{4}{c}{Optimal Transmit Powers \textbf{P}$^{\prime}$}(watts) & \multicolumn{3}{c}{Number of Iterations}\cr
    \cmidrule(lr){2-5} \cmidrule(lr){6-8}
    &$P_1^{\prime}$&$P_2^{\prime}$&$P_3^{\prime}$&$P_4^{\prime}$ &Proposed Scheme & max-utility Scheme & max-SINR Scheme\cr
    \midrule
    $0.05$ &2.500&2.4054&2.4144&2.500&12&7&N/A \cr
    $0.10$ &2.500&2.3840&2.3748&2.500&23&N/A &N/A \cr
    $0.15$ &2.500&2.3761&2.3731&2.500&34&N/A &N/A \cr
    $0.20$ &2.500&2.3699&2.3699&2.500&51&N/A &N/A \cr
    \bottomrule
    \end{tabular}
    \end{threeparttable}
\end{table*}

Fig. \ref{fig:CapVSWv} to Fig. \ref{fig:CapVSDens} show that the proposed scheme outperforms the two comparisons in terms of network throughput. In addition to replacing the equality allocation with on-demand spectrum allocating among AVs, the performance improving is also due to the transmit power controlling in the proposed scheme. Taking scenarios with four different AV densities, i.e., $0.05\,$AV/m, $0.10\,$AV/m, $0.15\,$AV/m, and $0.20\,$AV/m, as examples, the optimal transmit powers obtained by the proposed scheme are shown in Table \ref{tab:performance_comparison}. To avoid the impact of the initial APs' transmit powers on the network throughput, APs' transmit powers are fixed on $2.5\,$watts with communication range of $260\,$m for both comparisons. For the scenario with $0.05\,$AV/m AV density, the network throughputs achieved by the proposed, the max-utility, and the max-SINR schemes are $0.86\,$Gbits/s, $0.52\,$Gbits/s, and $1.12\,$Gbits/s, respectively. However, both of the max-utility and the max-SINR schemes are ineffective to scenarios with $0.10\,$AV/m, $0.15\,$AV/m, and $0.20\,$AV/m, due to the high inter-cell interferences. From columns $2$ to $5$ in Table \ref{tab:performance_comparison}, the transmit powers of AP $2$ and AP $3$ for the proposed scheme have been adjusted, which helps control inter-cell interference for both eNBs and the other two APs' transmission. Despite the improvement in network throughput, the computational complexity of the proposed scheme is higher than the other two, resulting in more iterations, as shown in columns $6$ to $8$ of Table \ref{tab:performance_comparison}.

\begin{figure}[htbp]
\centering
\includegraphics[height=2.5 in]{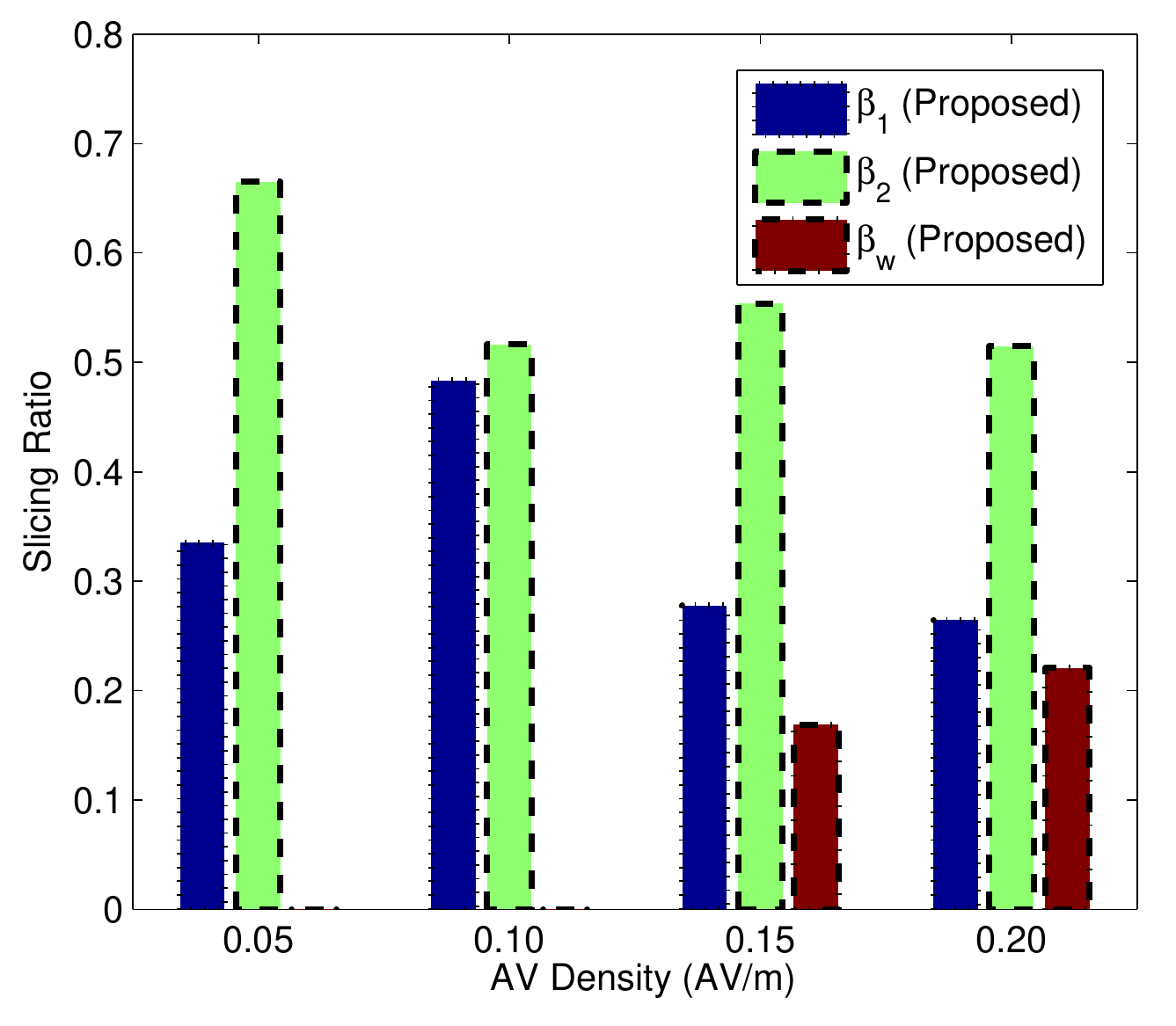}
\caption{Spectrum slicing ratios under different AV density for the proposed scheme with $p=0.8$.}
\label{fig:slicing}
\end{figure}

In addition to adjusting the transmit powers for the APs, the spectrum slicing ratios among BSs are also adjusted by the proposed scheme, as shown in Fig. \ref{fig:slicing}. With AV density increasing from $0.05\,$AV/m to $0.20\,$AV/m, the amount of spectrum resources sliced to Wi-Fi APs, i.e., the spectrum slicing ratio $\beta_w$, is increased in the proposed scheme. This is because a large $\beta_w$ indicates more spectrum resources can be reused among APs and therefore improving the spectrum efficiency. 


\section{Conclusions}
\label{sec:CONCLUSIONS}
In this paper, we have proposed a dynamic spectrum management framework to enhance spectrum resource utilization in the MEC-based AVNET with the consideration of cellular and Wi-Fi interworking. Through enabling NFV control modules at the MEC servers, the spectrum resource utilization can be enhanced via dynamically and centrally managing a wide range of spectrum resources and adjusting the transmit power for Wi-Fi APs. To maximize the aggregate network utility and provide QoS-guaranteed downlink transmission for delay-sensitive and delay-tolerant applications, three optimization problems have been investigated to slice spectrum among BSs fairly, to allocate spectrum among AVs associated with a BS in a QoS-guaranteed way, and to control transmit powers of Wi-Fi APs. 
In order to solve these three problems, 
we first use linear programming relaxation and first-order Taylor series approximation to transform them into tractable forms, and then design an ACS algorithm to jointly solve them. Based on the simulation results, the designed ACS algorithm has good convergence property within acceptable number of iterations. Compared with two existing spectrum management schemes, the proposed framework is more robust to AV density changing and provides higher network throughput.

\begin{appendices}
\section{Definitions and operations}
\label{Appendix:Def1}

\begin{myDef}
\textbf{Second-order conditions:} Suppose function $f$ is twice differentiable, i.e., it has Hessian or second derivative, $\nabla^2 f$, at each point in its domain, \textbf{dom}$f$. Then $f$ is concave if and only if \textbf{dom}$f$ is a convex set and its second derivative is negative semidefinite for all $y\in \textbf{dom} f$, i.e., $\nabla^2 f \preceq 0$.
\end{myDef}

To express biconcave set and biconcave function, we define $A\subseteq\mathbb{R}^n$ and $B\subseteq\mathbb{R}^m$ as two non-empty convex sets, and let $Y$ be the Cartesian product of $A$ and $B$, i.e., $Y\subseteq A\times B$. Define $a$- and $b$-sections of $Y$ as $Y_a =\{b\in B:(a,b)\in Y\}$ and $Y_b =\{a\in A:(a,b)\in Y\}$.

\begin{myDef}
\textbf{Biconcave set:} Set $Y\subseteq A\times B$ is called as a biconcave set on $A\times B$, if $Y_a$ is convex for every $a\in A$ and $Y_b$ is convex for every $b\in B$.
\end{myDef}

\begin{myDef}
\textbf{Biconcave function:} Define function $f$: $Y\rightarrow \mathbb{R}$ on a biconvex set $Y\subseteq A\times B$. Then function $f$: $Y\rightarrow \mathbb{R}$ is called a biconcave function on $Y$, if $f_a(b)=f(a,b):Y_a\rightarrow \mathbb{R}$ is a concave function on $Y_a$ for every given $a\in A$, and $f_b(a)=f(a,b):Y_b\rightarrow \mathbb{R}$ is a concave function on $Y_b$ for every given $b\in B$.
\end{myDef}

\begin{myDef}
\textbf{Biconcave optimization problem:} An optimization problem with form ${\rm max}\{f(a,b):(a,b)\in Y\}$ is called as a biconcave optimization problem, if the feasible set $Y$ is biconvex on $Y_a$ and $Y_b$, and the objective function $f(a,b)$ is biconcave on $Y$.
\end{myDef}

\begin{myOpe}
\textbf{Nonnegative weighted sums:} A nonnegative weighted sum of concave functions is concave.
\end{myOpe}

\begin{myOpe}
\textbf{Composition with an affine mapping:} Let function $\hbar:\mathbb{R}^n \rightarrow \mathbb{R}$, $E \in \mathbb{R}^{n\times m}$, and $e \in \mathbb{R}^n$. Define function $\ell:\mathbb{R}^m\rightarrow \mathbb{R}$ by $\ell(y) = \hbar(Ey + e)$ with \textbf{dom} $\ell = \{y | Ey + e \in$ \textbf{dom} $\hbar\}$. Then function $\ell$ is concave if $\hbar$ is concave.
\end{myOpe}

\section{Proof of Proposition 1}
\label{Appendix:Pro2}
\begin{proof}
For problem (P1$^\prime$), constraint (\ref{BRP1_1_Cons_1}) indicates that $\{\beta_1, \beta_2, \beta_w\}$ is a closed set, i.e., the problem domain is a convex set, and the objective function of (P1$^\prime$) is the summation of AVs' logarithmic utilities, where the logarithmic function is a concave function due to the non-positive second derivative. Moreover, for an AV associated to a BS, the utility is logarithm of the achievable transmission rate, and the corresponding achievable transmission rate is an affine function of $\beta_1$, $\beta_2$, or $\beta_w$. Thus, based on the above two operations, we can conclude that the objective function of problem (P1$^\prime$) is a concave function on the three optimal variables $\beta_1$, $\beta_2$, and $\beta_w$. Furthermore, constraint (\ref{BRP1_Cons_1}) can be rewritten into inequality concave constraints, such as $\beta_1 \in [0,1]$ can be as $-\beta_1 \leq 0$ and $\beta_1 \leq 1$, and constraint (\ref{BRP1_Cons_2}) is an equality affine function. Therefore, problem (P1$^\prime$) is a concave optimization problem.
\end{proof}

\section{Proof of Proposition 2}
\label{Appendix:Pro3}
\begin{proof}
Constraints (\ref{BRP2_1_Cons_13})-(\ref{BRP2_1_Cons_1}) of problem (P2$^\prime$) indicate that $\{\widetilde{\textbf{X}}, \widetilde{\textbf{X}}^\prime\}$ and $\{\textbf{R}, \textbf{R}^\prime\}$ are convex sets, and Cartesian product is an operation that preserves convexity of convex sets \cite{boyd2004convex}. Thus, the domain of (P2$^\prime$)', $\{\widetilde{\textbf{X}}, \widetilde{\textbf{X}}^\prime\}\times \{\textbf{R}, \textbf{R}^\prime\}$, is a convex set. Moreover, as stated before, the objective function of (P2$^\prime$) is the summation of AVs' achievable transmission rates from the associated BSs, where the transmission rate achieved by an AV $k$ is an affine function on elements of $\{\textbf{R}, \textbf{R}^\prime\}$ for a given association pattern and is an affine function on the association pattern variable for a given resource allocation. Considering the affine function is both concave and convex, it can prove that the objective function of problem (P2$^\prime$) is a biconcave function on variable set $\{\widetilde{\textbf{X}}, \widetilde{\textbf{X}}^\prime\}\times \{\textbf{R}, \textbf{R}^\prime\}$. Moreover, constraints (\ref{BRP2_1_Cons_2}) and (\ref{BRP2_1_Cons_3}) are equality affine on $\{\widetilde{\textbf{X}}, \widetilde{\textbf{X}}^\prime\}$, constraints (\ref{BRP2_1_Cons_13})-(\ref{BRP2_1_Cons_16}) are equality biaffine on $\{\widetilde{\textbf{X}}, \widetilde{\textbf{X}}^\prime\}\times \{\textbf{R}, \textbf{R}^\prime\}$, constraints (\ref{BRP2_1_Cons_12})-(\ref{BRP2_1_Cons_1}) are respectively inequality affine on $\{\widetilde{\textbf{X}}, \widetilde{\textbf{X}}^\prime\}$ and $\{\widetilde{\textbf{R}}, \widetilde{\textbf{R}}^\prime\}$, and constraints (\ref{BRP2_1_Cons_4})-(\ref{BRP2_1_Cons_11}) are inequality biaffine on $\{\widetilde{\textbf{X}}, \widetilde{\textbf{X}}^\prime\}\times \{\textbf{R}, \textbf{R}^\prime\}$. Thus, we can conclude that (P2$^\prime$) is a biconcave optimization problem.
\end{proof}

\end{appendices}

\bibliographystyle{IEEEtran}
\bibliography{HeterogeneousJournal}

\begin{thebibliography}{10}
\providecommand{\url}[1]{#1}
\csname url@samestyle\endcsname
\providecommand{\newblock}{\relax}
\providecommand{\bibinfo}[2]{#2}
\providecommand{\BIBentrySTDinterwordspacing}{\spaceskip=0pt\relax}
\providecommand{\BIBentryALTinterwordstretchfactor}{4}
\providecommand{\BIBentryALTinterwordspacing}{\spaceskip=\fontdimen2\font plus
\BIBentryALTinterwordstretchfactor\fontdimen3\font minus
  \fontdimen4\font\relax}
\providecommand{\BIBforeignlanguage}[2]{{%
\expandafter\ifx\csname l@#1\endcsname\relax
\typeout{** WARNING: IEEEtran.bst: No hyphenation pattern has been}%
\typeout{** loaded for the language `#1'. Using the pattern for}%
\typeout{** the default language instead.}%
\else
\language=\csname l@#1\endcsname
\fi
#2}}
\providecommand{\BIBdecl}{\relax}
\BIBdecl

\bibitem{hussain2018autonomous}
R.~Hussain and S.~Zeadally, ``Autonomous cars: Research results, issues and
  future challenges,'' \emph{IEEE Commun. Surv. Tutor.}, to appear.

\bibitem{human2018li}
L.~Li, K.~Ota, and M.~Dong, ``Human-like driving: Empirical decision-making
  system for autonomous vehicles,'' \emph{IEEE Trans. Veh. Technol.}, to
  appear.

\bibitem{su2018distributed}
Z.~Su, Y.~Hui, and T.~Luan, ``Distributed task allocation to enable
  collaborative autonomous driving with network softwarization,'' \emph{IEEE J.
  Sel. Areas Commun.}, to appear.

\bibitem{li2018consensus}
Y.~Li, C.~Tang, K.~Li, X.~He, S.~Peeta, and Y.~Wang, ``Consensus-based
  cooperative control for multi-platoon under the connected vehicles
  environment,'' \emph{IEEE Trans. Intell. Transp. Syst.}, no.~99, pp. 1--10,
  Sep. 2018.

\bibitem{peng2017performance}
H.~Peng, D.~Li, K.~Abboud, H.~Zhou, H.~Zhao, W.~Zhuang, and X.~Shen,
  ``Performance analysis of {IEEE 802.11 p DCF} for multiplatooning
  communications with autonomous vehicles,'' \emph{IEEE Trans. Veh. Technol.},
  vol.~66, no.~3, pp. 2485--2498, Mar. 2017.

\bibitem{peng2017resource}
H.~Peng, D.~Li, Q.~Ye, K.~Abboud, H.~Zhao, W.~Zhuang, and X.~Shen, ``Resource
  allocation for cellular-based inter-vehicle communications in autonomous
  multiplatoons,'' \emph{IEEE Trans. Veh. Technol.}, vol.~66, no.~12, pp.
  11\,249--11\,263, Dec. 2017.

\bibitem{peng2018vehicular}
H.~Peng, L.~Liang, X.~Shen, and G.~Y. Li, ``Vehicular communications: {A}
  network layer perspective,'' \emph{IEEE Trans. Veh. Technol.}, to appear.

\bibitem{sabuau2017optimal}
{\c{S}}.~Sab{\u{a}}u, C.~Oar{\u{a}}, S.~Warnick, and A.~Jadbabaie, ``Optimal
  distributed control for platooning via sparse coprime factorizations,''
  \emph{IEEE Trans. Auto. Control}, vol.~62, no.~1, pp. 305--320, Jan. 2017.

\bibitem{software2018zhang}
Y.~Zhang, H.~Zhang, K.~Long, Q.~Zheng, and X.~Xie, ``Software-defined and
  fog-computing-based next generation vehicular networks,'' \emph{IEEE Commun.
  Mag.}, vol.~56, no.~9, pp. 34--41, Sep. 2018.

\bibitem{yuan2018toward}
Q.~Yuan, H.~Zhou, J.~Li, Z.~Liu, F.~Yang, and X.~S. Shen, ``Toward efficient
  content delivery for automated driving services: {An} edge computing
  solution,'' \emph{IEEE Netw.}, vol.~32, no.~1, pp. 80--86, Jan. 2018.

\bibitem{hui2018content}
Y.~Hui, Z.~Su, T.~H. Luan, and J.~Cai, ``Content in motion: An edge computing
  based relay scheme for content dissemination in urban vehicular networks,''
  \emph{IEEE Trans. Intell. Transp. Syst.}, to appear.

\bibitem{hou2016vehicular}
X.~Hou, Y.~Li, M.~Chen, D.~Wu, D.~Jin, and S.~Chen, ``Vehicular fog computing:
  A viewpoint of vehicles as the infrastructures,'' \emph{IEEE Trans. Veh.
  Technol.}, vol.~65, no.~6, pp. 3860--3873, Feb. 2016.

\bibitem{feng2017ave}
J.~Feng, Z.~Liu, C.~Wu, and Y.~Ji, ``Ave: Autonomous vehicular edge computing
  framework with {ACO-based} scheduling,'' \emph{IEEE Trans. Veh. Technol.},
  vol.~66, no.~12, pp. 10\,660--10\,675, Jun. 2017.

\bibitem{abboud2016interworking}
K.~Abboud, H.~A. Omar, and W.~Zhuang, ``Interworking of {DSRC} and cellular
  network technologies for {V2X} communications: A survey,'' \emph{IEEE Trans.
  Veh. Technol.}, vol.~65, no.~12, pp. 9457--9470, Dec. 2016.

\bibitem{ETSI2018V2X}
\BIBentryALTinterwordspacing
``{ETSI GR MEC 022 V2.1.1} multi-access edge computing {(MEC)}; study on {MEC}
  support for {V2X} use cases,'' European Telecommunications Standards
  Institute, Tech. Rep., Sep. 2018. [Online]. Available:
  \url{https://www.etsi.org/deliver/etsi_gr/MEC/001_099/022/02.01.01_60/gr_MEC022v020101p.pdf}
\BIBentrySTDinterwordspacing

\bibitem{hu2018vehicular}
Q.~Hu, C.~Wu, X.~Zhao, X.~Chen, Y.~Ji, and T.~Yoshinaga, ``Vehicular
  multi-access edge computing with licensed {Sub-6 GHz, IEEE 802.11 p and
  mmWave},'' \emph{IEEE Access}, vol.~6, pp. 1995--2004, Dec. 2017.

\bibitem{Peng2018}
\BIBentryALTinterwordspacing
H.~Peng, Q.~Ye, and X.~Shen, ``{SDN}-based resource management for autonomous
  vehicular networks: {A} multi-access edge computing approach,'' \emph{arXiv
  preprint arXiv:1809.08966}, 2018. [Online]. Available:
  \url{https://arxiv.org/abs/1809.08966}
\BIBentrySTDinterwordspacing

\bibitem{quan2018software}
W.~Quan, K.~Wang, Y.~Liu, N.~Cheng, H.~Zhang, and X.~S. Shen,
  ``Software-defined collaborative offloading for heterogeneous vehicular
  networks,'' \emph{Wireless Commun. Mobile Comput.}, vol. 2018, pp. 1--9, Apr.
  2018.

\bibitem{herrera2016resource}
J.~G. Herrera and J.~F. Botero, ``Resource allocation in {NFV: A} comprehensive
  survey,'' \emph{IEEE Trans. Netw. Serv. Man.}, vol.~13, no.~3, pp. 518--532,
  Sep. 2016.

\bibitem{luo2018sdnmac}
G.~Luo, J.~Li, L.~Zhang, Q.~Yuan, Z.~Liu, and F.~Yang, ``sdnmac: A
  software-defined network inspired mac protocol for cooperative safety in
  {VANETs},'' \emph{IEEE Trans. Intell. Transp. Syst.}, vol.~19, no.~6, pp.
  2011--2024, Jun. 2018.

\bibitem{secinti2017software}
G.~Secinti, B.~Canberk, T.~Q. Duong, and L.~Shu, ``Software defined
  architecture for {VANET: A} testbed implementation with wireless access
  management,'' \emph{IEEE Commun. Mag.}, vol.~55, no.~7, pp. 135--141, Jul.
  2017.

\bibitem{liang2017vehicular}
L.~Liang, H.~Peng, G.~Y. Li, and X.~Shen, ``Vehicular communications: A
  physical layer perspective,'' \emph{IEEE Trans. Veh. Technol.}, vol.~66,
  no.~12, pp. 10\,647--10\,659, Dec. 2017.

\bibitem{Ye2018}
Q.~Ye, W.~Zhuang, S.~Zhang, A.~Jin, X.~Shen, and X.~Li, ``Dynamic radio
  resource slicing for a two-tier heterogeneous wireless network,'' \emph{IEEE
  Trans. Veh. Tech.}, to appear.

\bibitem{ye2013user}
Q.~Ye, B.~Rong, Y.~Chen, M.~Al-Shalash, C.~Caramanis, and J.~G. Andrews, ``User
  association for load balancing in heterogeneous cellular networks,''
  \emph{IEEE Trans. Wireless Commun.}, vol.~12, no.~6, pp. 2706--2716, Jun.
  2013.

\bibitem{liang2016virtual}
C.~Liang, F.~R. Yu, H.~Yao, and Z.~Han, ``Virtual resource allocation in
  information-centric wireless networks with virtualization,'' \emph{IEEE
  Trans. Veh. Tech.}, vol.~65, no.~12, pp. 9902--9914, Dec. 2016.

\bibitem{boyd2004convex}
S.~Boyd and L.~Vandenberghe, \emph{Convex optimization}.\hskip 1em plus 0.5em
  minus 0.4em\relax Cambridge university press, 2004.

\bibitem{gorski2007biconvex}
J.~Gorski, F.~Pfeuffer, and K.~Klamroth, ``Biconvex sets and optimization with
  biconvex functions: a survey and extensions,'' \emph{Math. method oper.
  res.}, vol.~66, no.~3, pp. 373--407, Jun. 2007.

\bibitem{guo2018fast}
C.~Guo, C.~Shen, Q.~Li, J.~Tan, S.~Liu, X.~Kan, and Z.~Liu, ``A fast-converging
  iterative method based on weighted feedback for multi-distance phase
  retrieval,'' \emph{Sci. rep-UK}, vol.~8, pp. 1--10, Apr. 2018.

\bibitem{lema2017business}
M.~A. Lema, A.~Laya, T.~Mahmoodi, M.~Cuevas, J.~Sachs, J.~Markendahl, and
  M.~Dohler, ``Business case and technology analysis for {5G} low latency
  applications,'' \emph{IEEE Access}, vol.~5, pp. 5917--5935, Apr. 2017.

\end{thebibliography}

\ifCLASSOPTIONcaptionsoff
  \newpage
\fi

\end{document}